\documentclass{aa}
\usepackage{graphicx}
\usepackage{amsmath}
\usepackage{txfonts}
\usepackage{lscape}
\usepackage{gensymb}
\usepackage{color}
\usepackage[hyperindex,breaklinks=true, colorlinks, citecolor=blue, draft=True]{hyperref}
\usepackage{natbib}

\usepackage[switch,pagewise]{lineno}

\graphicspath{{figures/}}
\makeatletter

\DeclareRobustCommand{\ion}[2]{\textup{#1\,\textsc{\lowercase{#2}}}}
\DeclareRobustCommand{\kms}{km\,${\rm s}^{-1}$}
\DeclareRobustCommand{\jyb}{Jy\,beam${}^{-1}$}

\begin{document} 


\title{Feedback in W49A diagnosed with radio recombination lines and models}

 \author{
        M. R. Rugel\inst{1}\and
        D. Rahner\inst{2}\and
        H. Beuther\inst{1}\and
        E. W. Pellegrini\inst{2}\and
        Y. Wang\inst{1}\and
        J. D. Soler\inst{1}\and
        J. Ott\inst{3}\and
        A. Brunthaler\inst{4}\and
        L. D. Anderson\inst{5,6,7}\and
        J. C. Mottram\inst{1}\and
        T. Henning\inst{1}\and
        \mbox{P. F. Goldsmith\inst{8}}\and
        M. Heyer\inst{9}\and
        R. S. Klessen\inst{2,10}\and
        S. Bihr\inst{1}\and
        K. M. Menten\inst{4}\and
        R. J. Smith\inst{11} \and
        J. S. Urquhart\inst{12}\and 
        S. E. Ragan\inst{13}\and
        \mbox{S. C. O. Glover\inst{2}}\and
        N. M. McClure-Griffiths\inst{14}\and
        F. Bigiel\inst{15,2} \and
        N. Roy\inst{16}
        }

  \institute{
        Max Planck Institute for Astronomy, K\"onigstuhl 17, 69117 Heidelberg, Germany\\ \email{rugel@mpia.de} \and
        Universit\"at Heidelberg, Zentrum f\"ur Astronomie, Institut f\"ur
        Theoretische Astrophysik, Albert-Ueberle-Str. 2, 69120
        Heidelberg, Germany \and
        National Radio Astronomy Observatory, PO Box O, 1003 Lopezville Road, Socorro, NM 87801, USA \and
        Max Planck Institute for Radioastronomy, Auf dem H\"ugel 69, 53121 Bonn, Germany\and
        Department of Physics and Astronomy, West Virginia University, Morgantown WV 26506, USA \and
        Adjunct Astronomer at the Green Bank Observatory, P.O. Box 2, Green Bank WV 24944, USA \and
        Center for Gravitational Waves and Cosmology, West Virginia University, Chestnut Ridge Research Building, Morgantown, WV 26505, USA \and
        Jet Propulsion Laboratory, California Institute of Technology, 4800 Oak Grove Drive, Pasadena, CA 91109, USA \and
        Department of Astronomy, University of Massachusetts, Amherst, MA 01003, USA \and
        Universit\"{a}t Heidelberg, Interdisziplin\"{a}res Zentrum f\"{u}r Wissenschaftliches Rechnen, INF 205, 69120 Heidelberg, Germany\and
        Jodrell Bank Centre for Astrophysics, School of Physics and Astronomy, 
        The University of Manchester, Oxford Road, Manchester, M13 9PL, UK\and
        School of Physical Sciences, University of Kent, Ingram Building, Canterbury, Kent CT2 7NH, UK \and
        School of Physics and Astronomy, Cardiff University, Queen's Buildings, The Parade, Cardiff, CF24 3AA, UK\and
        Research School of Astronomy and Astrophysics, The Australian National University, Canberra, ACT, Australia\and
        Argelander-Institut f\"ur Astronomie, Universit\"at Bonn, Auf dem H\"ugel 71, 53121 Bonn, Germany\and
        Department of Physics, Indian Institute of Science, Bangalore 560012, India
        }

\abstract
{
We present images of radio recombination lines (RRLs) at wavelengths around 17\,cm from the star-forming region W49A to determine the kinematics of ionized gas in the THOR survey (The \ion{H}{i}/OH/Recombination line survey of the inner Milky Way) at an angular resolution of $16\farcs8\times13\farcs8$. 
The distribution of ionized gas appears to be affected by feedback processes from the star clusters in W49A. 
The velocity structure of the RRLs shows a complex behavior with respect to the molecular gas. 
We find a shell-like distribution of ionized gas as traced by RRL emission surrounding the central cluster of OB stars in W49A.
We describe the evolution of the shell with the recent feedback model code {\sc warpfield} that includes the important physical processes and has previously been applied to the 30 Doradus region in the Large Magellanic Cloud.
The cloud structure and dynamics of W49A are in agreement with a feedback-driven shell that is re-collapsing. 
The shell may have triggered star formation in other parts of W49A. 
We suggest that W49A is a potential candidate for star formation regulated by feedback-driven and re-collapsing shells. 
}

\maketitle

\section{Introduction} \label{sec:introduction}
Radiative and mechanical feedback from UV radiation and winds from OB stars and supernova explosions alter the potential of molecular gas to fragment and form future generations of stars, but their precise role in star formation is yet to be clarified.
Recent investigations of 30 Doradus in the Large Magellanic Cloud (LMC) indicate that feedback from the older stellar population 
in its main star cluster NGC 2070 has been unable to destroy its parent molecular cloud \citep{RahnerPellegrini:2018aa}. 
Rather, feedback-driven shells of ionized and molecular gas from these older cluster members may have re-collapsed due to the gravitational attraction 
of the star cluster and the self-gravity of the shell, forming a second generation of stars. 
Only with this second generation can the combined stellar feedback become strong enough to disperse the parent molecular cloud.

While star formation induced by re-collapsing shells may not change the total number of stars formed in a giant molecular cloud (GMC), 
such a process would naturally imply that star formation in at least some GMCs may occur episodically. 
As pointed out by \citet{RahnerPellegrini:2017aa}, re-collapse of a feedback-driven shell can occur only in the most massive and densest GMCs. 
It is therefore an open question whether this also takes place in GMCs in the Milky Way.

In our search for a 30 Doradus analog in the Milky Way, we focus this study on W49A.
It is one of the most massive and luminous young star-forming regions in the Galaxy, with one of the highest luminosity-to-molecular-mass ratios, indicating a high star formation efficiency \citep[it is responsible for 12\% of all luminosity in the Galaxy;][]{UrquhartKonig:2018aa}. 
While regions like W49A lie within the statistical distribution of luminosity-to-molecular-mass ratios of the Galaxy-wide sample of star-forming regions, 
the nature of the physical processes that cause this extreme star formation is still not fully understood. 
This makes it an ideal region to search for re-collapse of feedback-driven shells.

\begin{figure*}
  \includegraphics[width=0.99\textwidth]{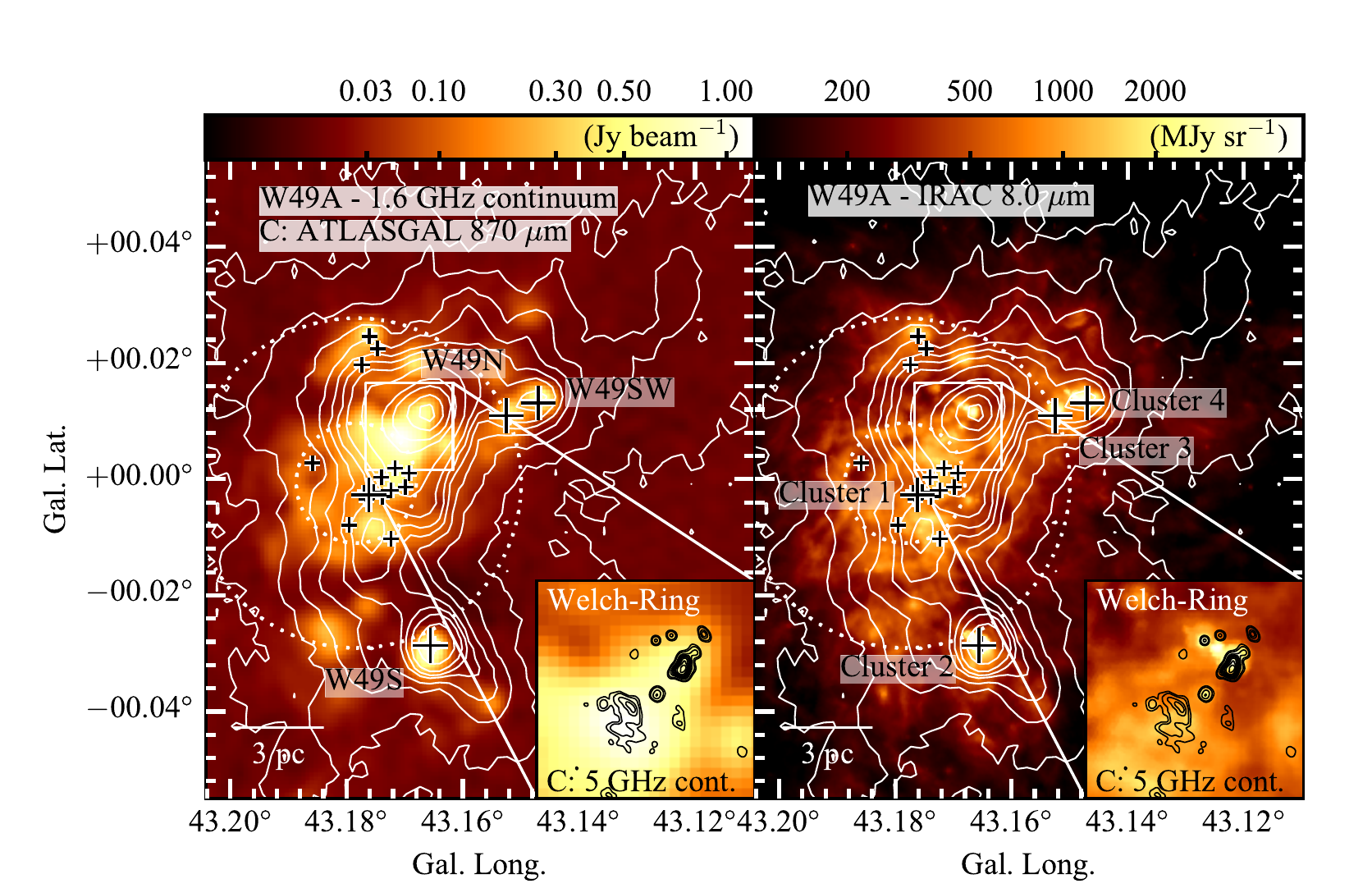}
  \caption{Overview of W49A in 1.6 GHz continuum emission (in color scale on the {\it left panel}; THOR; \citealt{WangBihr:2018aa}) 
                and GLIMPSE Spitzer IRAC 8.0 $\mu$m emission ({\it right panel}; \citealt{BenjaminChurchwell:2003aa,ChurchwellBabler:2009aa}). 
                 White contours indicate cold, dense dust in 870\,$\mu$m emission (ATLASGAL, \citealt{SchullerMenten:2009aa}; 
                at levels of -0.24, 0.24, 0.8, 1.6, 2.4, 3.2, 4.0, 5.2, 8.0,16.0, 24.0, 40.0, 56.0\,\jyb). 
                {Large crosses} indicate stellar clusters found with JHK imaging \citep{AlvesHomeier:2003aa} while small crosses indicate individual O stars observed by \citet{WuBik:2016aa}. The insert shows the
                Welch-Ring of ultra-compact \ion{H}{ii} regions (UC\,\ion{H}{ii} regions, \citet{WelchDreher:1987aa}), 
                represented here by contours of 5 GHz continuum emission (CORNISH survey, \citet{HoarePurcell:2012aa}; 
                at levels of 0.01, 0.02, 0.05, 0.75, 0.1, 0.125, 0.15, 0.175, 0.2, 0.25\,\jyb).
                The  dotted inner ring indicates the shell structure found in this work 
                (in good agreement with shells found by \citealt{PengWyrowski:2010aa}); 
                the  outer ring indicates the approximate size of W49A.
                Labels in the left panel indicate main cm-continuum emission peaks \citep[e.g.,][]{DreherJohnston:1984aa} 
                and in the right panel indicate infrared star clusters \citep{AlvesHomeier:2003aa}.}
  \label{fig:0}
\end{figure*}

W49 was first discovered as a radio source by  \citet{Westerhout:1958aa} and is located at a distance of 11.1\,kpc from the Sun \citep{ZhangReid:2013aa} in the Perseus spiral arm \citep[e.g.,][]{MooreUrquhart:2012aa}, at a  Galactocentric radius similar to that of the solar system.
It is associated with a GMC of mass $\sim 10^6\,{\rm M_\odot}$ \citep{Galvan-MadridLiu:2013aa}, which contains the massive \ion{H}{ii} region complex W49A \citep{MezgerSchraml:1967aa} with a molecular gas mass of $\sim2\times 10^5\,{\rm M_\odot}$ \citep{UrquhartKonig:2018aa,Galvan-MadridLiu:2013aa}. 
An overview of W49A in Galactic coordinates is shown in Fig.~\ref{fig:0}, with 8\,$\mu$m emission \citep[GLIMPSE; ][]{BenjaminChurchwell:2003aa,ChurchwellBabler:2009aa} as a tracer of ongoing star formation (e.g., \citealt{StockPeeters:2014aa}), and 870\,$\mu$m emission \citep[from ATLASGAL;][]{SchullerMenten:2009aa} indicating the presence of cold and dense gas -- the sites of future star formation. The main physical parameters are summarized in Table~\ref{tbl:1}.

The W49A cloud harbors multiple ultracompact (UC) \ion{H}{ii} regions \mbox{\citep[e.g.,][]{DreherJohnston:1984aa,De-PreeMehringer:1997aa}}, with a central condensation in a ring-like structure \citep{WelchDreher:1987aa}. The region has the highest concentration of UC\,\ion{H}{ii} regions in the Galactic disk \citep[18 compact and UC \ion{H}{ii} regions;][]{UrquhartThompson:2013aa}. Four star clusters are detected by infrared imaging \citep{AlvesHomeier:2003aa}, amounting to a total stellar mass of $5-7\times10^4\,{\rm M_\odot}$ \mbox{\citep{HomeierAlves:2005aa}}. The so-called `Cluster 1' \citep{HomeierAlves:2005aa} in the central part of W49A (r$\lesssim 2.5\,{\rm pc}$) contains about 50 O stars with $M_*\ge20\,{\rm M_\odot}$, and a total of $\sim$270 O stars are found in and around W49A \citep{HomeierAlves:2005aa}.
Several of these candidate cluster members have masses between $20$ and $250\,{\rm M_\odot}$, as determined from near-infrared spectroscopy and photometry \mbox{\citep[][Fig.~\ref{fig:a3}]{WuBik:2016aa}}. 

\begin{table*}
\caption{Cloud properties of W49A.}
\small
\begin{tabular}{p{4cm}|l|p{4cm}|p{5cm}}
\hline
\hline
Quantity             & Assumed values & Comments & Reference\\
\hline
Cloud radius                                                    & $R\approx6\,{\rm pc}$                                                   & Approximate distance from Cluster 1 to  W49SW and W49S. & \citet{PengWyrowski:2010aa} \\
Molecular gas mass (W49A; r$\lesssim6$ pc)& $M_{\rm gas}\approx2^{+2}_{-1}\times10^5\,{\rm M_\odot}$ & Assumed uncertainties of a factor of 2; used for modeling.&\citet{Galvan-MadridLiu:2013aa}, \citet{UrquhartKonig:2018aa}\\
Molecular gas mass (W49 molecular cloud; r$\lesssim60$ pc)                                                     & $M_{\rm gas}\approx1\times10^6\,{\rm M_\odot}$ &  & \citet{Galvan-MadridLiu:2013aa}\\
Stellar cluster mass (all infrared sub-clusters)& $M_*\approx5-7\times10^4\,{\rm M_\odot}$ &                                & \citet{HomeierAlves:2005aa}\\
Stellar cluster mass (Cluster 1)                      & $M_*\approx1\times10^4\,{\rm M_\odot}$    & Used for modeling, assuming that mass is possibly up to a factor of two higher. & \citet{HomeierAlves:2005aa}\\
Derived molecular gas density (W49A; \mbox{$r\lesssim6\,{\rm pc}$})   & $n\approx4_{-2}^{+4}\times10^{3}\,{\rm cm^{-3}}$ & Assuming homogeneity and spherical symmetry. & \\
\hline
\end{tabular}
\label{tbl:1}
\end{table*}

The molecular gas of W49A shows a complex velocity structure, with blue- and red-shifted components with respect to the systemic velocity ($\varv_{\rm LSR} = 8.6$\,\kms; \citealt{QuirezaRood:2006ab}): 
There are two main velocity components at 4\,\kms\ and 12\,\kms (or at $-$4.6\,\kms\, and 3.4\,\kms\ with respect to the systemic velocity) observed in many gas tracers on scales ranging from the entire W49A region \citep{MufsonLiszt:1977aa,MiyawakiHayashi:1986aa,SimonJackson:2001aa,MiyawakiHayashi:2009aa} to the innermost parts, towards W49N in CS \citep[e.g.,][]{SerabynGuesten:1993aa}.
While CO transitions may be optically thick, the double-peaked structure appears to reflect the true dynamics of the molecular cloud, as it is persistent in optically thin, high-density tracers such as H$^{13}$CO$^+$(1-0) and CS(2-1) (Fig. 3 in \citealt{Galvan-MadridLiu:2013aa}; also \citealt{MiyawakiHayashi:2009aa}). 
The velocity components were attributed to background and foreground clouds, respectively, by \citet{SerabynGuesten:1993aa}, as indicated by ${\rm H_2CO}$ absorption \citep{GossTilanus:1985aa}.
This spatial location would mean that the two clouds are moving towards each other, either as the collision of two clouds, or the inside-out collapse of one cloud \citep{SerabynGuesten:1993aa,WelchDreher:1987aa}. 
Alternatively, the velocity components could be due to expanding motions \citep{PengWyrowski:2010aa}. 
Other studies have suggested that the complexity of the ${\rm H_2CO}$ absorption, as well as of CS and HCO$^+$ emission in this region, is connected to a complex arrangement of \ion{H}{ii} regions surrounded by dense and diffuse molecular gas, outflows, and infall motions \citep{DickelGoss:1990aa,WilliamsDickel:2004aa}. 

The number of ionizing photons from the stellar population in W49A \citep{AlvesHomeier:2003aa}, with the largest contribution being from the most massive of these stars \citep{WuBik:2016aa}, is sufficient, or may exceed that required, to produce the observed ionized gas within W49A, as determined from low- \citep{Kennicutt:1984aa} and high-resolution continuum observations \citep[e.g.,][]{De-PreeMehringer:1997aa}. The fraction of ionized to molecular gas mass is low ($\sim1\%$; \citealt{Galvan-MadridLiu:2013aa}), indicating that the molecular gas in W49A is not yet penetrated by this radiation. Estimates of the radiation pressure on dust particles indicate that radiative feedback alone from the star cluster is not yet strong enough to disperse W49A \citep{Galvan-MadridLiu:2013aa,ReisslKlessen:2018aa}, while models with other feedback mechanisms (thermal pressure from the \ion{H}{ii} region and shocked winds,  protostellar jets, radiation pressure; but not wind-momentum) find feedback to be approximately strong enough to disrupt the cloud \citep{MurrayQuataert:2010aa}. It was indeed recently noted that comprehensive modeling that includes all sources of feedback is needed to reliably predict the impact of radiative feedback on cloud evolution \citep{RahnerPellegrini:2017aa}. 

This poses the question of  how, if at all, feedback from Cluster~1 affected star formation in W49A. 
The  \ion{H}{ii} regions and O/B associations within W49A have different ages, with some having already dispersed some of the molecular gas around them (the O/B stars discovered in \citealt{AlvesHomeier:2003aa}), and others still deeply embedded (e.g., the UC\,\ion{H}{ii} regions in the Welch ring; e.g., \citealt{DreherJohnston:1984aa,De-PreeMehringer:1997aa}). 
Both single O stars and UC\,\ion{H}{ii} regions appear to be spread across W49A in small groups.
While many studies focussed on the formation of the UC\,\ion{H}{ii} regions in the Welch-ring/W49N dust clumps \citep[e.g.,][]{WelchDreher:1987aa,SerabynGuesten:1993aa,WilliamsDickel:2004aa}, two main scenarios have been invoked to explain the star formation in W49A as a whole. 
One interpretation is causally unrelated, sub-clustered star formation across W49A \citep{AlvesHomeier:2003aa} along filamentary inflows of molecular gas from a larger reservoir of gas \citep{Galvan-MadridLiu:2013aa}. 
Alternatively, star formation and the dynamics of the region are causally connected to the feedback of the central star cluster. The central cluster of O stars drives expanding shells of molecular gas \citep{PengWyrowski:2010aa}, which may have triggered star formation in the Welch ring \citep{AlvesHomeier:2003aa}. Feedback may also be responsible for the formation of W49S and W49SW, if they are interpreted as ejecta from the central star cluster. 
While previous interpretations of sub-clustered star formation may be a suitable description of the star formation activity in W49A, the dynamics show imprints of feedback \citep{PengWyrowski:2010aa}. 

In this work, we investigate the observational signatures of stellar feedback using emission from radio recombination lines (RRLs).
These lines have been used in numerous works to determine the physical conditions of ionized gas in \ion{H}{ii} regions (e.g., \citealt{HjellmingDavies:1970aa}, \citealt{Shaver:1980aa}; see also review by \citealt{RoelfsemaGoss:1992aa}), in particular also for W49A \citep[e.g.,][]{van-GorkomGoss:1980aa,BalserBania:1999aa}.
They have also been mapped for different star-forming regions in the Milky Way \citep[e.g.,][]{PankoninWalmsley:1979aa,LangGoss:2001aa,BalserGoss:2001aa}. 
In this work, we use high-$n$ RRLs (1.6-1.9\,GHz) to trace the kinematics of ionized gas in W49A. The continuum emission at these frequencies may be optically thick, especially towards the location of UC\,\ion{H}{ii} regions such as the Welch-Ring. As the continuum optical depth affects the line emission, this caveat is addressed and the results are compared to previous RRL studies at higher frequencies where available. 
We use the RRL data from the THOR survey (The \ion{H}{i}/OH/Recombination line survey of the inner Milky Way; \citealt{BeutherBihr:2016aa}) as constraints for models of the past and future evolution of a feedback-driven shell in a cloud with average physical properties corresponding to those of W49A. 

After presenting the observations in Sect.~\ref{sec:observationsandmethods}, 
the RRL emission is analyzed and compared to molecular gas emission in Sect.~\ref{sec:results}
to characterize the morphology and velocity structure of different gas phases in W49A. 
We identify shell-like RRL emission at the interface between hot ionized gas and neutral gas.
With the radius and velocity of the shell, we aim to constrain its evolution with the 1D code, {\sc warpfield}\footnote{\url{https://bitbucket.org/drahner/warpfield/}} \citep{RahnerPellegrini:2017aa} in Sect.~\ref{sec:models}. While acknowledging that a 1D code does not capture all the complex dynamics of W49A, it includes a refined treatment of the interplay of the relevant types of stellar feedback and allows us to easily probe a large parameter space in density and star formation efficiency. This provides general conclusions on the evolution of this feedback-driven shell in W49A.
The results are discussed in the context of the previous interpretations of star formation in W49A, 
and compared to the feedback-driven star formation history of 30 Doradus in Section~\ref{sec:discussion}. In Sect.~\ref{sec:conclusion}, we summarize our main findings.

\section{Observations and Methods} \label{sec:observationsandmethods}

\subsection{Radio recombination lines at 1.6--1.9 GHz}

The RRLs allow us to study the kinematics and structure of the ionized gas in W49A. We choose RRLs between orders of n=151 and n=158 (1.6-1.9 GHz) from the THOR survey (The \ion{H}{i}/OH/Recombination line survey of the inner Milky Way; \citealt{BeutherBihr:2016aa}) to extract spatially resolved kinematic information of ionized gas. As these lines may in principle be affected by high optical depth in the continuum, we discuss the optical depth in W49A in Sect.~\ref{sec:rrlemission} and compare our data in Sect.~\ref{sec:results:vpeak} to kinematic information obtained from RRLs at higher frequencies in the literature. 

The H151$\alpha$-H156$\alpha$ and H158$\alpha$ RRLs were observed towards W49A within the THOR survey. Observations of 5$-$6 minutes per pointing were conducted with the VLA C-configuration in L-band. Each line was observed with a bandwidth of 2\,MHz and a spectral resolution of 15.63\,kHz. 
This corresponds to a velocity resolution of 2.5\,\kms\ and 2.8\,\kms\ for the highest and lowest frequency transitions, respectively (the H151$\alpha$ line at 1.891\,GHz and the H158$\alpha$ line at 1.651\,GHz; see also Table~2 in \citealt{BeutherBihr:2016aa}). 
Using {\sc CASA}\footnote{\url{http://casa.nrao.edu}; version 4.2.2}, the absolute flux scale and the bandpass were calibrated on the quasar 3C386 and the complex gain with the quasar J1925+2106.

With a significant detection of all RRLs, the data were gridded to a spectral resolution of 5\,\kms\ as opposed to 10\,\kms\ in the first data release of the THOR survey. 
All observations were continuum-subtracted, imaged, and combined into mosaics of $3.75\degree\times2.5\degree$, and deconvolved as described in \citet{BeutherBihr:2016aa}. 
We only consider the seven lines in the frequency interval between 1.6 and 1.9 GHz to maintain higher spatial resolution (while higher-order transitions from lower frequencies were included by \citealt{BeutherBihr:2016aa}). 
The angular resolution of the observations of the individual transitions is between $12\farcs0\times14\farcs6$ (H151$\alpha$) and $16\farcs8\times13\farcs8$ (H158$\alpha$). 
To improve the signal-to-noise ratio, the different Hn$\alpha$ lines were stacked in velocity after regridding each individual line to a common angular resolution (using {\sc CASA} task {\tt imsmooth}). 
Our image of the stacked continuum-subtracted emission of the RRLs has an angular resolution of $16\farcs8\times13\farcs8$. 
The RMS noise of the stacked image in a line-free channel is 1.5\,m\jyb. The characteristics of the RRL data are summarized in Table~\ref{tbl:2}.

\begin{table*}
\caption{Datasets used in this work.}
\small
\begin{tabular}{p{4.0cm}|rrrrr|l}
\hline
\hline
Data & Frequency & Angular res. & Spectral res. & Noise & Survey & Reference\\
 & [GHz] & & [\kms] &  & \\
\hline
Stacked RRLs (H151$\alpha$-H156$\alpha$; H158$\alpha$) & 1.780$^{a}$ & $16\farcs8\times13\farcs8$ & 5 & $\sigma(F)\sim1.5$\,m\jyb & THOR$^{b}$ & This work \\
$^{13}$CO(3-2) & 330.587 &  $15\arcsec$ & 0.5 & $\sigma(T_A)\sim0.6\,{\rm K}$ & CHIMPS$^{c}$ & \citet{RigbyMoore:2016aa} \\
C$^{18}$O(3-2) & 329.331 &  $15\arcsec$ & 0.5 & $\sigma(T_A)\sim0.8\,{\rm K}$ & CHIMPS$^{c}$ & \citet{RigbyMoore:2016aa} \\
$^{13}$CO(1-0) & 110.201 &  $46\arcsec$ & 0.2& $\sigma(T_A)\sim0.1\,{\rm K}$ & GRS$^{d}$ & \citet{JacksonRathborne:2006aa} \\
\hline
\end{tabular}
\tablefoot{$(a):$ Mean frequency of all lines; $(b):$ The \ion{H}{i}/OH/Recombination line survey of the inner Milky Way; $(c):$ The \mbox{$^{13}$CO/C$^{18}$O (J = 3$\rightarrow$2)} Heterodyne Inner Milky Way Plane Survey; $(d):$ The Galactic Ring Survey.}
\label{tbl:2}
\end{table*}

\begin{figure}
  \includegraphics[width=0.99\columnwidth]{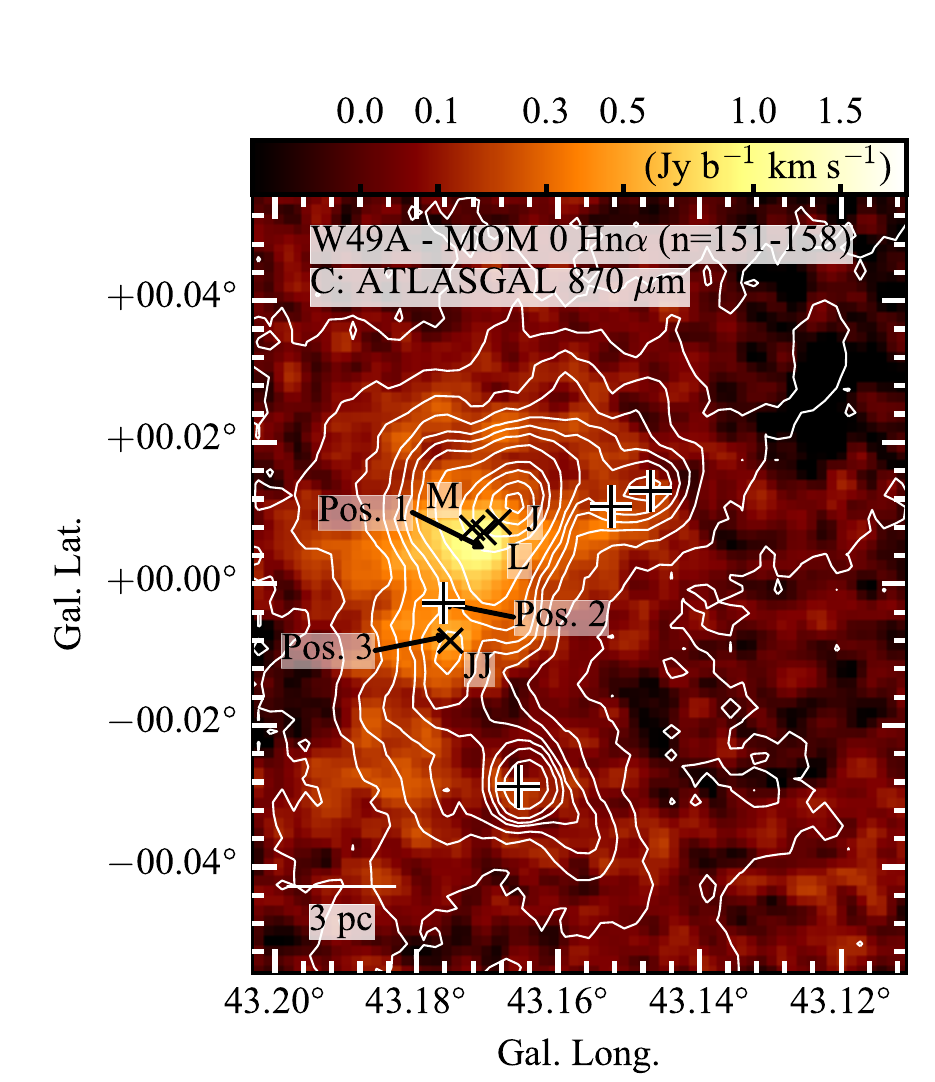}
  \caption{Integrated emission of stacked Hn$\alpha$ (n=151-158; angular resolution $16\farcs8\times13\farcs8$, 0.85 pc). 
                The color scale shows velocity integrated intensity (between $-$40 and 60\,\kms).
                Contours ({\it white}) show 870\,$\mu$m emission (ATLASGAL survey; \citealt{SchullerMenten:2009aa}; 
                contours as in Fig.~\ref{fig:0}). {Crosses} indicate infrared star clusters \citep{AlvesHomeier:2003aa}. 
                The positions of the spectra shown in Fig.~\ref{fig:2} are marked, as well as the locations of the 3.6\,cm continuum sources ``M'', ``L'', ``J'' and ``JJ'' \citep{De-PreeMehringer:1997aa}, which are mentioned in Sects.~\ref{sec:rrlemission}~and~\ref{sec:discussion:dynamics} ({\it ``x''-shaped symbols}).}
  \label{fig:1}
\end{figure}

\begin{figure}
  \includegraphics[height=0.85\textheight]{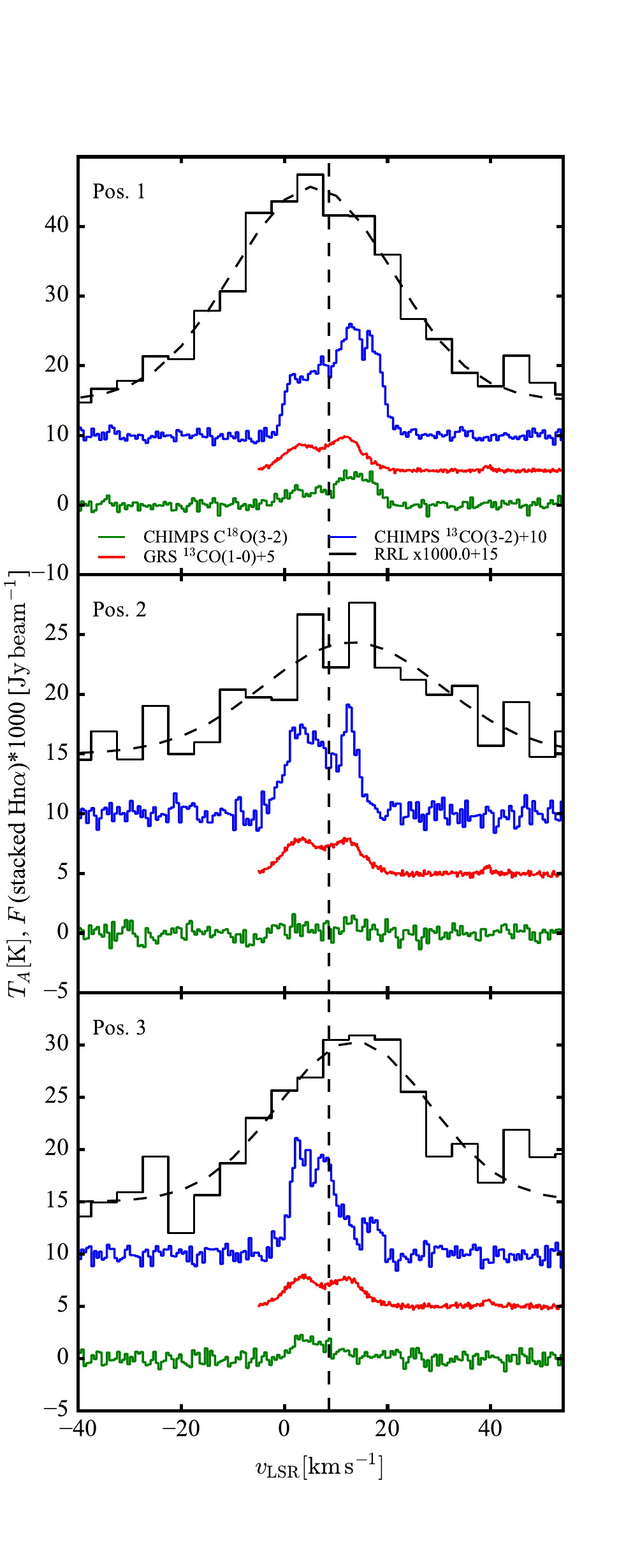}
  \caption{Spectra of Hn$\alpha$ (n=151-158; {\it black}), CHIMPS ${}^{13}$CO(3-2) ({\it blue}),  GRS ${}^{13}$CO(1-0) ({\it red}; data extends only to $-$5\,\kms) and CHIMPS C${}^{18}$O(3-2) ({\it green}), at selected locations in W49A, as indicated in Figs.~\ref{fig:1}~and~\ref{fig:3}. The RRL emission is scaled by a factor of 1000, and given in \jyb, while the antenna temperature ($T_A$) of the CO emission is given in K. The black dashed lines indicate the Gaussian fits to the RRL emission mentioned in Sect.~\ref{sec:gaussian_decomposition}. The vertical black dashed line denotes the systemic velocity ($\varv_{\rm LSR} = 8.6$\,\kms; \citealt{QuirezaRood:2006ab}). The channel spacing is 5\,\kms for the RRLs, 0.5\,\kms\ for CO spectra from CHIMPS, and 0.21\,\kms\ for GRS.}
  \label{fig:2}
\end{figure}

\begin{figure*}
  \includegraphics[width=0.99\textwidth]{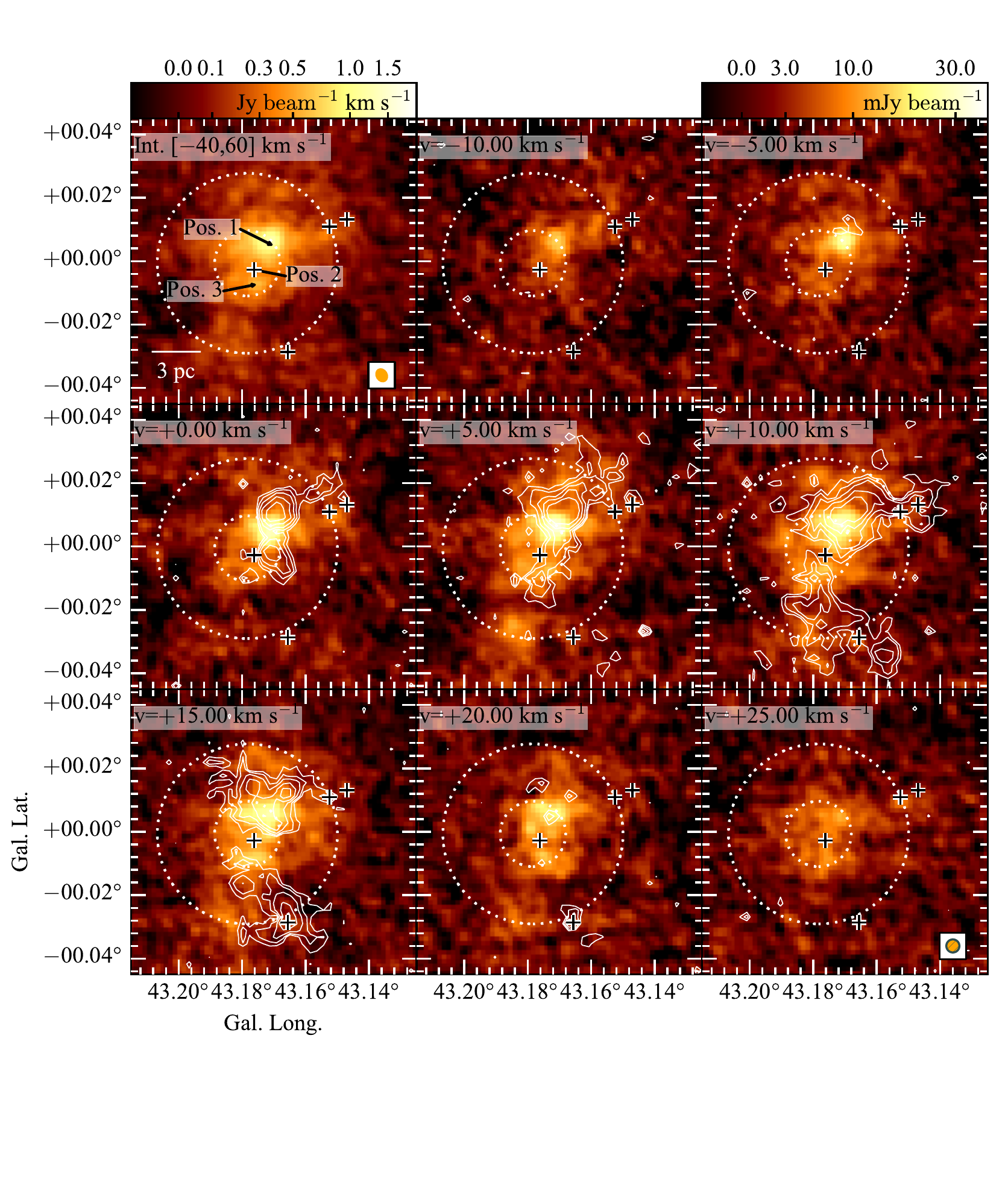}
  \caption{Channel maps of stacked Hn$\alpha$ n=151-158 shown in color scale. The upper left panel shows the moment 0 image (velocity integrated emission; Fig.~\ref{fig:1}), with the positions and numbers of the spectra in Fig.~\ref{fig:2} overlaid. The remaining panels show the RRL emission between $-$10 and 25\,\kms, with the corresponding contours of ${\rm C{}^{18}O}$ emission (CHIMPS; \citealt{RigbyMoore:2016aa}; at levels of $T_A$ =  0.65, 1.03, 1.63, 2.59, 4.1 K) after binning to 5\,\kms\ channels. 
  The white dashed circles denote the shell of an expanding \ion{H}{ii} region (inner ring) and the assumed extent of W49A (outer ring) as described in Fig.~\ref{fig:0}. The center of both radii is $l=43.1783$\degree, $b=-0.0007$\degree. Crosses denote stellar clusters identified in \citet{AlvesHomeier:2003aa}. The angular resolution of the RRL emission is shown in the lower-right corner of the left panel in the top row (\textit{orange}). The angular resolution of the ${\rm C{}^{18}O}$ emission is shown in the right panel of the bottom row, superposed as an empty hatched circle in gray on the RRL beam.
  }
  \label{fig:3}
\end{figure*}

This study focusses on the continuum-subtracted RRL emission associated with W49A ($\varv_{\rm LSR} = 8.6$\,\kms; \citealt{QuirezaRood:2006ab}). A likely unassociated component at 60\,\kms, which has been found in other RRL observations (e.g., \citealt{LiuMcIntyre:2013aa}; Liu et al., in prep.), is not significantly detected in our dataset and not discussed further here (faint emission (<3\,$\sigma$) at 60\,\kms\ is seen in some of the spectra in Fig.~\ref{fig:2}).
The integrated emission between $-$40 and 60\,\kms\ (moment 0) is presented in Fig.~\ref{fig:1}. 
Individual spectra are shown in Fig.~\ref{fig:2}, which were extracted towards three positions indicated by the labels ``Pos.~1'', ``Pos.~2'' and ``Pos.~3'' in Fig.~\ref{fig:1} and in the upper-left panel of Fig.~\ref{fig:3}. 
Images of all channels between $-$10 and 25\,\kms\ are shown in Fig.~\ref{fig:3}. 

\subsection{Archival CO observations}
To investigate the interplay between ionized and molecular gas, we use $^{13}$CO(3-2) and C$^{18}$O(3-2) emission (CHIMPS\footnote{The \mbox{$^{13}$CO/C$^{18}$O (J = 3$\rightarrow$2)} Heterodyne Inner Milky Way Plane Survey.} survey; \citealt{RigbyMoore:2016aa}) to trace the morphology and kinematics of molecular gas in W49A. 
Although $^{13}$CO may be optically thick at the high column densities found towards W49A \citep{Galvan-MadridLiu:2013aa}, 
lines from high-density tracers like CS and H$^{13}$CO$^+$, as well as from the high-column-density tracer C$^{18}$O, show similar shapes in the same study. 
The ratio of main-beam brightness temperatures of the two CO isotopologs remains close to constant at $\sim0.11$ across the peaks and troughs of the spectra, roughly in agreement with a ratio of $0.14$ from \citet{WilsonRood:1994aa} in the solar neighborhood. 
This is expected as W49A has a similar Galactocentric radius to that of the Sun. 
Similar conclusions have been obtained by \citet{MiyawakiHayashi:2009aa} based on investigations of \mbox{$^{13}$CO (J=1-0)} and \mbox{C$^{18}$O (J=1-0)} at 17\arcsec\ resolution. 
Hence, we assume $^{13}$CO emission to be optically thin and use it with the \mbox{C$^{18}$O(3-2)} emission for the kinematic analysis of the molecular gas in W49A.

Table~\ref{tbl:2} contains a detailed description of the CO data used, which includes archival $^{13}$CO(1-0) data from the GRS\footnote{Galactic Ring Survey.} survey \citep[][]{JacksonRathborne:2006aa} at lower angular resolution.
Spectra of all tracers are shown in Fig.~\ref{fig:2} towards selected positions in W49A. 
Figure~\ref{fig:3} shows emission contours of C$^{18}$O(3-2) after smoothing to a spectral resolution of 5\,\kms. 

\subsection{Gaussian line-fitting of the RRL data} \label{sec:gaussian_decomposition}
\begin{figure*}
  \includegraphics[width=0.99\textwidth]{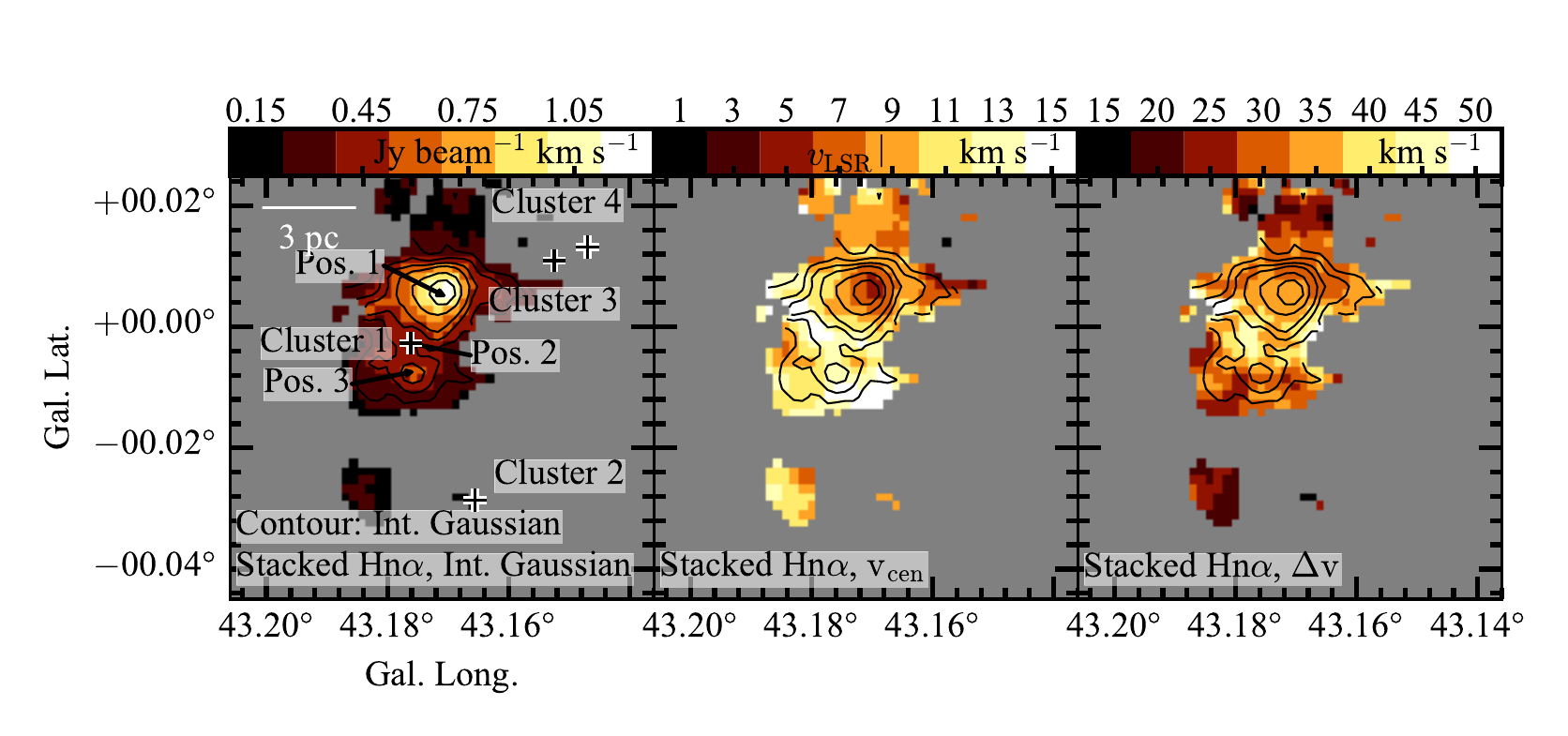}
  \caption{Integrated emission ({\it left panel}), peak velocity ({\it middle panel}), and FWHM ({\it right panel}) of the  
                Gaussian-decomposed, stacked Hn$\alpha$ (n=151--158) emission. Crosses indicate infrared star clusters \citep{AlvesHomeier:2003aa}. The positions of the spectra shown in Fig.~\ref{fig:2} are marked. The systemic velocity is marked in the color-bar of the middle panel ($\varv_{\rm LSR} = 8.6$\,\kms; \citealt{QuirezaRood:2006ab}).
                The contours represent integrated emission of the fitted Gaussian (shown in color in the left panel; at levels of 0.3, 0.4, 0.5, 0.6, 0.8, 1.0 and 1.2 \jyb\,\kms).}
  \label{fig:5}
\end{figure*}

To extract the kinematic properties of the RRL emission, maps of velocity centroids and line widths are derived for each tracer.
The RRLs typically have a full width at half maximum (FWHM) of 25--45\,\kms\ in W49A. 
The RRL data are well described by a single Gaussian. 
We fit the stacked RRLs at each pixel with a Gaussian profile with the {\sc CASA} task {\tt specfit} to obtain maps of the peak velocity and FWHM (Fig.~\ref{fig:5}).
Despite the large channel width of 5\,\kms, the nominal uncertainty on the center of the Gaussian is $\lesssim2.5$\,\kms\ in all cases, and $<1$\,\kms\ towards the RRL emission peak. 

\section{Results} \label{sec:results}
\subsection{Spatial distribution of the RRL emission in comparison to other gas tracers} \label{sec:rrlemission}
The emission in the stacked Hn$\alpha$ data between 1.6 and 1.9 GHz is spread over the entire W49A star-forming complex. 
The spectral indices of the continuum emission between 1 and 2 GHz ($\alpha$) fall between $\sim0$ and $\sim1$, indicating thermal, partially optically thick emission \citep[][figure 10]{WangBihr:2018aa}. Towards W49N, the spectral index approaches values of $\alpha\sim0.8$, which can indicate ionized outflows \citep[e.g.,][and references therein]{De-PreeMehringer:1997aa}.
It is similarly extended in W49A as the 8\,$\mu$m continuum (Fig.~\ref{fig:0}) and the 870\,$\mu$m continuum emission (Fig.~\ref{fig:1}).

The emission peaks of the different tracers are slightly different. 
The integrated 1.6--1.9-GHz RRLs peak offset from the 870\,$\mu$m emission (Fig.~\ref{fig:1}).
The offset between the peaks is $\sim$25\arcsec\ or $\sim$1.5\,pc. 

The peak in 870\,$\mu$m emission corresponds to W49N, 
which harbors the Welch-Ring of UC\,\ion{H}{ii} regions \citep{WelchDreher:1987aa}, as indicated in Fig.~\ref{fig:0} by the 5-GHz emission. 
The fact that the 1.6--1.9-GHz RRL emission (and the continuum emission at the same frequency) peaks slightly offset is likely due to optically thick continuum from the high-density ionized gas in the very compact UC\,\ion{H}{ii} regions at the location of the Welch-ring that lowers the flux at our longer wavelengths.
A similar effect could be causing the relatively weak RRL emission towards W49S and W49SW, 
which also host UC\,\ion{H}{ii} regions \citep[e.g.,][]{DreherJohnston:1984aa,De-PreeMehringer:1997aa}.

The peak of the 1.6--1.9-GHz RRL emission occurs towards the more extended radio sources ``L'' and ``M'' at the edge of the Welch ring \citep[e.g.,][see Fig.~\ref{fig:1} and inlay to Fig.~\ref{fig:0}]{DreherJohnston:1984aa}.
Since it is also offset from the dust emission peak, this \ion{H}{ii} region may have already cleared parts of its cocoon of gas and dust. 

\subsection{Shell-like distribution of RRL and CO emission around Cluster 1} \label{sec:results:shell}
At the center of W49A, the emission of stacked RRLs appears to have a shell-like morphology.
The channel maps (Fig.~\ref{fig:3}) show that towards the position of Cluster~1 (Pos.~2 in Fig.~\ref{fig:2}), 
the RRL emission is suppressed. 
North and south of Cluster~1 (with respect to Galactic coordinates), the RRL emission increases, in parts resembling a ring-like structure.
This emission is potentially indicative of an ionized bubble around Cluster~1 with a radius of $\sim2\pm1$\,pc, 
as indicated by the inner white ring in Fig.~\ref{fig:3}. 

The radius of the bubble is in agreement with the shells found in 4.5 and 8.0\,$\mu$m emission by \citet[][2-3 pc diameter]{PengWyrowski:2010aa};
the RRL emission here agrees better with their shell~1, though this cannot be said with certainty due to the lower angular resolution of the RRL image. 
Extended emission at this position is seen in the broadband 1.6~GHz continuum image (Fig.~\ref{fig:0}) as well as in radio continuum at higher frequencies \citep[e.g.,][]{De-PreeMehringer:1997aa}. The spectral index between 1 and 2\,GHz in this particular region of W49A lies between $\alpha = -0.1$ and 0.3, and increases to $\alpha=0.8$ towards the Welch-Ring \citep[][figure 10]{WangBihr:2018aa}. 

Confirmation of this structure is seen in the \mbox{C$^{18}$O(3-2)} data (Fig.~\ref{fig:3}), but it is less pronounced in \mbox{$^{13}$CO(3-2)} and (1-0) emission (Fig.~\ref{fig:4}). 
In Fig.~\ref{fig:3} at 5\,\kms, the C$^{18}$O emission resembles an arc around Cluster~1. 
Channel maps at 10\,\kms\ and 15\,\kms show elongated emission edges around Cluster~1, together with a lack of emission towards the center of the bubble (see also spectrum of Pos.~2 in Fig.~\ref{fig:2}). 
To a lesser degree, this can also be seen in \mbox{$^{13}$CO(3-2)} and (1-0) emission in Fig.~\ref{fig:4}. At 10\,\kms\ and 15\,\kms\ the emission is lower towards Cluster~1 than to the north or south of it. 

We do not observe a full ring in RRL or CO emission. Parts of the RRL emission may be due to outflows from the surrounding UC\,\ion{H}{ii} regions. RRL emission appears towards the north, south, and west of Cluster~1 only (in the Galactic coordinate frame). The CO observations show similar structure. 
Together with arcs in 8~$\mu$m emission reported east of Cluster~1 by \citet[][see also Fig.~\ref{fig:0}]{PengWyrowski:2010aa}, 
we attribute the RRL emission to a ring-like structure indicative of a bubble around Cluster~1. 

\subsection{Kinematics of RRLs and $^{13}$CO} \label{sec:results:kinematics}
\subsubsection{Velocity distribution} \label{sec:results:vpeak}
The \mbox{$^{13}$CO(3-2)} emission shows two components at similar strength towards Cluster 1 (Pos.~2). 
Blue-shifted with respect to the systemic velocity of W49A ($\varv_{\rm LSR} = 8.6$\,\kms; \citealt{QuirezaRood:2006ab}) lies a broad component with a peak at $\sim$4\,\kms\ (or multiple blended narrow components between 1$-$7\,\kms). 
The other component is a narrow, red-shifted emission peak at $\sim$13\,\kms. 
The RRL emission at this position is located between the CO components, although the peak occurs closer towards the red-shifted CO component, with a fitted peak of 13.6$\pm$2.7\,\kms, which has high uncertainties due to the weak RRL emission at Pos.~2. 

Across W49A, the stacked RRL emission shows variations in peak velocity from the north to the south. 
The peak velocity distribution of the RRL data (Fig.~\ref{fig:5}) indicates a peak velocity at the latitude of W49N (Pos.~1 and higher) of 6.2$\pm$0.7\,\kms.
At lower latitudes towards Pos.~3, velocities of 12.6$\pm$1.2\,\kms \ are observed. 
While RRL emission at low frequencies of $1-2$\,GHz may in principle be affected by optically thick continuum emission, we find that for large parts of the bubble, the spectral index is between $-$0.1 and 0.3, which is only slightly higher than what would be expected for optically thin emission. 
The velocities of the H92$\alpha$ RRL from \citet{De-PreeMehringer:1997aa} are comparable to the velocities of the higher-order RRLs shown here in the case of the sources ``JJ'' and Pos.~3 (14.3$\pm$0.3\,\kms\ and 12.6$\pm$1.2\,\kms, respectively). The velocity is lower for sources  ``M'' and ``L'' (3.8$\pm$0.4\,\kms\ and 1.8$\pm$0.13\,\kms) as compared to the high-order RRL emission in Pos.~1 (6.2$\pm$0.7\,\kms), which, however, may also be influenced by source ``J'', for which the  H92$\alpha$ RRL has a velocity of 9.3$\pm$0.5\,\kms.
The closeness of the RRL velocities from this work to those of RRL observations at higher frequencies indicates that potentially high optical depth in the continuum at 1-2\,GHz is not the governing factor determining the appearance of the kinematics. 

The \mbox{$^{13}$CO(3-2)} emission extends between 0 and 20\,\kms\ in W49A. 
Towards Pos.~1, a red-shifted peak
dominates the emission, while towards Pos.~3 it is a blue-shifted peak that  dominates. 
In both positions, the $^{13}$CO peak velocity appears anti-correlated with the peak velocity observed in the RRL emission at the same positions. 
At the RRL peak velocities, the $^{13}$CO emission seems to be suppressed.
The \mbox{C$^{18}$O(3-2)} emission shows similar profiles and confirms the location of the $^{13}$CO peaks. 
This confirms the assumption that the $^{13}$CO emission profiles are not affected by optical depth effects. 
The lower-angular-resolution GRS \mbox{$^{13}$CO(1-0)} emission shows two smooth peaks towards all three positions,
which are mainly the result of averaging over variations within the 46\arcsec\ GRS beam. 

To summarize, towards the position of Cluster~1, the RRL emission is located in between, or aligned with one of, the two $^{13}$CO emission components. 
The two $^{13}$CO components are separated by $\sim9$\,\kms. 
If these are attributed to fore- and background parts of the shell bubble \citep{PengWyrowski:2010aa}, the expansion or collapse velocity of this shell would be approximately between 5 and 10\,\kms, depending on whether both parts of the shell are moving, or one of them is at rest with respect to the star cluster.
An alternative explanation for the kinematics of the region is, for example, ionized outflows from the surrounding UC\,\ion{H}{ii} regions. 
To the north and south of the center of the shell bubble \citep{PengWyrowski:2010aa}, both the RRL and CO emission show one main component, with the peaks at each position anti-correlated with respect to the rest velocity of W49A ($\varv_{\rm LSR} = 8.6$\,\kms; \citealt{QuirezaRood:2006ab}). 
This indicates that, at least in parts of the ring, the emission may be influenced by dynamics of individual places of star formation in the region.
For the following modelling (Sect.~\ref{sec:models}) we concentrate on the large-scale dynamics. For the ionized bubble, we assume a spherical structure with a shell radius of 1$-$3\,pc from the RRL emission data and a relative velocity of the shell components of 5$-$10\,\kms. 

\subsubsection{Distribution of the FWHM of the RRL emission} \label{sec:results:fwhm}
It is intrinsically difficult to see expansion or infall signatures in the RRL line profiles, due to their large line widths of up to 50\,\kms. 
The large line width itself, especially towards Cluster 1 (note the increase of line widths up to $45$\,\kms\ in Fig.~\ref{fig:5}), however, may point to unresolved dynamical motions. 
Contributions to the line width are thermal broadening, pressure broadening, and nonthermal motions (dynamical broadening). The contribution of pressure broadening may be significant for high-order RRLs, while spatially unresolved dynamical motions can also contribute to the line width, especially in very active star-forming regions. 

We evaluate the relative importance of pressure and dynamical broadening for two positions. Assuming an electron temperature for W49A of $T_e\sim 8300\,{\rm K}$ \citep{QuirezaRood:2006aa}, 
the thermal line width is approximately 20\,\kms. 
An estimation of the pressure broadening requires observations of lower-order RRLs, 
as pressure broadening depends on the order of the energy level of the transition (n) to the seventh power \citep[e.g.,][]{KetoWelch:1995aa}. 
Observations of millimeter(mm)-H$\alpha$ lines (n=$39-42$) are available at an angular resolution of $\sim20\arcsec$ towards the ATLASGAL dust emission peaks in W49A \citep{KimWyrowski:2017aa}. 
For W49N (source AGAL043.166+00.011), a line width of 37.0$\pm$0.2\,\kms\ is reported for the mm-RRLs, while the centimeter(cm)-RRLs presented here have a line width of 31.4$\pm$3.3\,\kms. Since the widths are larger or comparable at most, we conclude that dynamical broadening dominates the emission, which is expected due to the large number of UC\,\ion{H}{ii} regions at this location. For this particular position, optical depth effects may also influence the kinematics of the cm-RRLs. Towards Pos.~3 (source AGAL043.178-00.011), the line width of the cm-RRLs is larger than the mm-RRLs (33.5$\pm$4.0\,\kms\ and 29.7$\pm$0.5\,\kms, respectively). Using equations 4 and 5 from \citet{Nguyen-LuongAnderson:2017aa} and a thermal line width of 20\,\kms, we derive line widths of dynamical and pressure broadening of 23\,\kms\ and 6\,\kms, respectively. Again, this indicates that dynamical broadening dominates the RRL line width over pressure broadening. 
While this is likely to be true for parts of W49A which contain high star formation activity (as UC\,\ion{H}{ii} regions indicate in many positions), pressure broadening is certainly present and may be dominant towards some positions. 
Especially for large line widths in Fig.~\ref{fig:5} (e.g., Pos.~2), observations of RRLs with lower quantum numbers at the same angular resolution as the cm-RRLs are necessary to verify the role of pressure broadening \citep[see, e.g.,][]{Nguyen-LuongAnderson:2017aa}. 

An investigation of the spectra for double-peaked structures, which would be indicative for relative motions of the front and back of a shell, is inconclusive: 
Figure~\ref{fig:3} shows slightly more emission in Pos.~2 at 5\,\kms\ and 15\,\kms, compared to 10\,\kms. 
The variation, however, is only at a 2-$\sigma$ level and the spectrum is not fitted well by two Gaussians. 
Also, two distinct Gaussian components with such a small velocity separation are not expected to be distinguishable in the RRL data from a single Gaussian due to their large intrinsic, thermal line widths. 

\section{Stellar feedback models for W49A} \label{sec:models} 
How has star formation progressed in W49A? We now use the presented data to answer important questions about W49A: What influence has the formation of high-mass stars had on the subsequent evolution of the cluster? To explore possible scenarios of the evolution of W49A, we employ models of an expanding shell around a star cluster to compare with our observations.

In the following, we compare our data to models of an expanding shell of a star cluster inside a molecular cloud to constrain the evolution of the shell around Cluster~1, which was discussed in Section~\ref{sec:results:shell}.
Given its complexity in terms of currently ongoing star formation and density distribution, we do not attempt here to reproduce the RRL emission in W49A directly. 
Rather we aim to use the radius at which the RRL emission peaks to constrain the evolutionary history of the shell.
As discussed below, we use literature values of the stellar mass in Cluster~1, the molecular cloud mass, and the average density of W49A (see Table~\ref{tbl:1}) as initial conditions, and constrain the evolutionary stage of the shell with its observed radius and the age of the stellar cluster. 

\citet{RahnerPellegrini:2017aa} have developed feedback models ({\sc warpfield}\footnote{The modelling in this paper is based on {\sc warpfield} 1 \citep{RahnerPellegrini:2017aa}. The recent update of the code ({\sc warpfield} 2; \citealt{RahnerPellegrini:2018ac}) was tested to yield consistent results for this analysis.}) describing the expansion of a hydrostatic, spherical shell due to radiative feedback from young stars, including the energy and momentum input from stellar winds and supernovae, while also accounting for the effects of the gravity of the cluster and the self-gravity of the cloud. 
Models with the physical prescription for the hydrostatic shell, which is used in {\sc warpfield}, have been successfully compared to observations in previous works \citep[e.g., M17;][]{PellegriniBaldwin:2007aa}. 
In the models by \citet{RahnerPellegrini:2017aa}, the expansion of the shell is initially governed by adiabatic expansion due to wind-shocked gas from the stellar winds of O/B stars in the massive star cluster. 
Once cooling becomes comparable to the wind-energy input, a second phase starts, and momentum-driven expansion from radiation and winds dominates. 
A potential third phase of expansion is reached as the entire molecular cloud is swept up and the system freely expands into the (low-density) ambient medium. 

These models are applied here due to their advantage in efficiently probing large parameter spaces in molecular cloud mass, density, and star formation efficiency.  
For the computation of multiple models to remain computationally feasible, the models assume a 1D geometry (i.e., spherical symmetry of the \ion{H}{ii} region). 
While this is certainly a limitation, especially for determining detailed dynamics and morphologies of \ion{H}{ii} regions, it allows us to trace the general evolution of feedback-driven shells. 
The models also incorporate all relevant physical aspects of feedback, in particular for very massive and luminous regions in which the initial mass function should be fully sampled, with masses of individual stars of up to 120\,${\rm M}_{\odot}$. 
In the sub-region of W49A relevant for this investigation, Cluster 1, the most massive star has a mass of $M=130\pm30\,{\rm M}_{\odot}$ \citep[][Fig.~\ref{fig:a3}]{WuBik:2016aa}, which is within the uncertainties of the highest mass of sampled stars in the models. 
Therefore, since W49A is one of the most massive and luminous regions in the Milky Way, it is an ideal target for such an investigation \citep{UrquhartKonig:2018aa}.

\subsection{Initial conditions} \label{sec:results:initialconditions}
The input parameters for the models are listed in Table~\ref{tbl:1}.
\citet{Galvan-MadridLiu:2013aa} find a molecular gas mass of $M_{\rm gas} \approx 2\times10^5\,{\rm M_\odot}$ for W49A within a radius of 6\,pc, which we adopt for the modeling in the following.
The value is in agreement with the cloud mass derived from dust emission \citep{SieversMezger:1991aa,UrquhartKonig:2018aa}.
As a side note, W49A is embedded in a larger molecular cloud complex, which has a total mass (including W49A) of  $1\times10^6\,{\rm M_\odot}$ within 60\,pc \citep{Galvan-MadridLiu:2013aa}. 
Hence, free expansion of the feedback-driven shell into the low-density ISM would only occur far beyond a radius of 6\,pc.
Since the shells in the models considered here do not expand beyond this point during the first expansion (see Sect.~\ref{sec:results:models}), we treat the molecular cloud mass as fixed at $M_{\rm gas} \approx 2\times10^5\,{\rm M_\odot}$ in the models. 
Low-density channels, in which free expansion would occur earlier, could exist in the true initial density distribution. Such channels could influence the expansion of the shell in the energy-driven phase, i.e., in the initial stages of the expansion. The momentum-driven expansion phase of the shell is not significantly affected. 

Assuming a homogenous medium and spherical geometry for $r<6$\,pc, the average initial molecular density of W49A is estimated as $n \sim  4\times10^3\,{\rm cm^{-3}}$. 
To address uncertainties of at least a factor of two in the molecular gas mass, 
multiple models are computed to explore a parameter space in density\footnote{Mass densities chosen as input to {\sc warpfield} are: $\rho\approx(8-40)\times10^{-21}\,{\rm g\,cm^{-3}}$, in steps of $4\times10^{-21}\,{\rm g\,cm^{-3}}$.} of $n =  (2-10)\times10^3\,{\rm cm^{-3}}$, in steps of $1\times10^3\,{\rm cm^{-3}}$.
The initial radial gas density profile of the models is assumed to be flat for simplicity. Implications of radially decreasing density profiles and inhomogeneities in general are discussed in Sects.~\ref{sec:results:models},~\ref{sec:discussion:models},~and~\ref{sec:discussion:future}.
It must be noted that in reality, the cloud density might evolve as a function of time due to inflow of gas onto the cloud \citep[e.g.,][]{FukuiKawamura:2009aa,SeifriedWalch:2017aa,Ibanez-MejiaMac-Low:2017aa}. This is currently not taken into account in {\sc warpfield}.

With literature constraints on the mass of the stellar clusters in W49A \citep{AlvesHomeier:2003aa} and the above assumption on the molecular cloud mass, 
we constrain the star formation efficiency (SFE).
The mass of the central O/B-star cluster, Cluster 1 (Fig.~\ref{fig:0}), is  $M_* = 1\times10^4\,{\rm M_\odot}$ \citep{HomeierAlves:2005aa}, 
but may potentially be higher due to the large visual extinction in the region. 
Therefore, we choose SFEs of $\epsilon= 0.05$ and 0.09 to account for star cluster masses of $M_* \simeq (1-2)\times10^4\,{\rm M_\odot}$.

For these parameter ranges in density and SFEs the cluster does not disperse its cloud in the first 2 Myr, assuming a cloud mass of $M_{\rm cl} = 2\times10^5\,{\rm M_\odot}$ (see Table~\ref{tbl:1}; \citealt{Galvan-MadridLiu:2013aa,UrquhartKonig:2018aa}), 
nor does it drive feedback shells beyond the outer radii of W49A (see Sect.~\ref{sec:results:models}). 
The total cloud mass is only of dynamical importance once free expansion into the ambient ISM sets in. 
Since this stage is not reached in any of the models in the first 2 Myr, 
the important parameters are the density and SFE. 
We choose not to vary the initial molecular cloud mass of $M_{\rm cl} = 2\times10^5\,{\rm M_\odot}$, which is the mass in W49A within $r<6\,{\rm pc}$. 
The shell will encounter the ambient ISM only at $r>60\,{\rm pc}$, 
since W49A is embedded in a larger molecular cloud complex. 
Therefore, the assumption of a fixed initial cloud mass is reasonable, as the phase of free expansion of the shell is likely to start only at larger radii, and after accumulating more mass than assumed here.

\subsection{Models of stellar feedback indicate collapse of an initial expanding shell} \label{sec:results:models}
\begin{figure*}
  \includegraphics[width=0.99\textwidth]{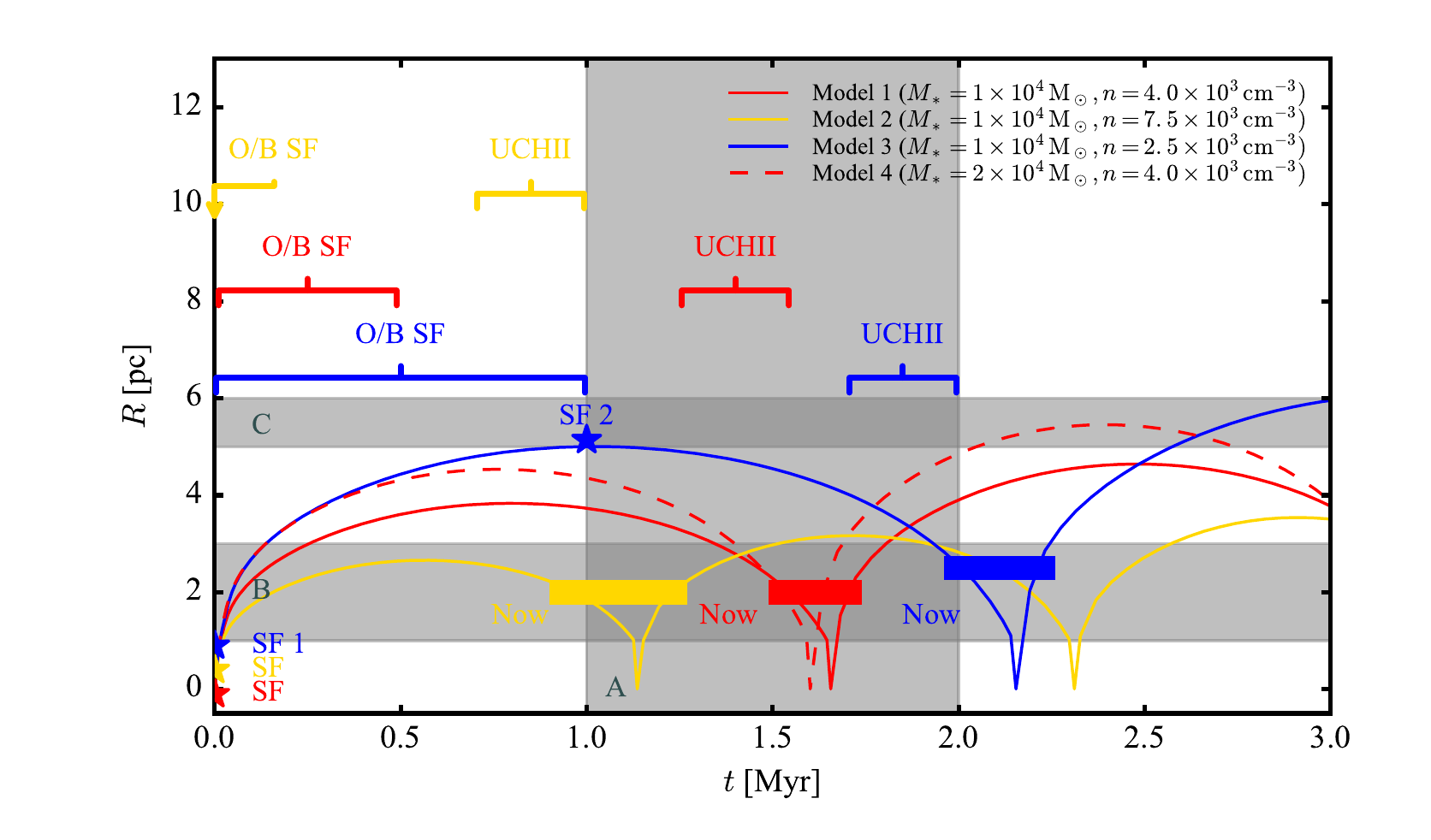}
  \caption{Expansion radius vs. time for models of stellar cluster feedback ({\sc warpfield}; Rahner et al., 2017) in W49A. $t=0$ denotes the formation of the first stellar cluster. 
  Observational constraints on stellar cluster age ($A$), shell size ($B$) and extent of W49A ($C$) are shown in gray.  
  Models~1-3 were computed with molecular gas densities of $n=(4.0, 7.5, 2.5)\times10^3\,
{\rm cm^{-3}}$, respectively, and a stellar cluster mass of $M_* = 1\times10^4\,{\rm M_\odot}$. Model~4 used a molecular gas density of $n=4.0\times10^3\,
{\rm cm^{-3}}$ and a stellar cluster mass of $M_* = 2\times10^4\,{\rm M_\odot}$. For simplicity, the molecular gas density was assumed to be constant (see text for further discussion). 
  Each model agrees best with the observations at the expansion radius indicated by a {\it filled bar} (``Now''), 
  which then defines the elapsed time. 
  The {\it stars} with labels ``SF'' or ``SF 1'' denote the time of the formation of the first stellar cluster in the model. 
  To put the models into perspective, the observational constraints from the perspective of today ({\it filled bars}) 
  on the age of the O/B cluster and the UC\,\ion{H}{ii} regions are highlighted by brackets in the top part of the plot (``O/B SF'' or ``UC\,\ion{H}{II}''). 
  The observational constraints on the O/B star formation in Model 2 (indicated by a {\it yellow arrow}) coincides 
  approximately with the formation of the first star cluster in the model.
  For Model 3, ``SF 2'' marks the radius and time at which the model reaches the outskirts of W49A 
  (no additional star formation is added to the model at this point). 
  }
  \label{fig:6}
\end{figure*}

The time evolution of the shell radius in each model is shown in Fig.~\ref{fig:6}, with $t=0$ at the beginning of the expansion of the shell. 
For simplicity, we neglect the (ultra)-compact \ion{H}{ii}-region phase of the high-mass stars, which is assumed to occur before the shell expansion and is considerably smaller than the stellar ages ($3\times10^5$\,yr; \citealt{MottramHoare:2011aa}).
The shell is expected to have a similar age to the O stars, that is, 1-2\,Myr \citep{WuBik:2016aa}. 
This time range is highlighted in Fig.~\ref{fig:6} in gray, 
together with the adopted current expansion radius of 1-3 pc (in comparison, the angular resolution of the RRL images is $\sim0.8$\,pc), 
in agreement with both the RRL emission (see Sect.~\ref{sec:results:shell}) and the 8~$\mu$m emission \citep{PengWyrowski:2010aa}.
For reference, the approximate radius of the entire W49A region is shown as the gray-shaded region C in Fig.~\ref{fig:6}. 

These conditions are met by several models: 
For a large range of densities, shells fall within the limits of the radius after an evolution of 1-2 Myr. 
In all models in Fig.~\ref{fig:6}, the observed shell size agrees with the simulated shell size at two distinct times -- once during collapse and again during re-expansion.
For three models ($n=2.5, 4.0, 7.5\times10^{3}\,{\rm cm}^{-3}$),
the radius and range in time at which the observational constraints are met are indicated schematically in Fig.~\ref{fig:6} (by a colored bar which is labeled ``now'').
If the double-peaked velocity profile in the $^{13}$CO data is attributed to relative motions of the back and the front of the shell, it would be expanding or collapsing at a speed of 5$-$10\,\kms\ (see Sect.~\ref{sec:results:kinematics}). 
This agrees well with the modeled velocities of the re-collapsing shells. 
For each model, the formation period of the O/B-stars is highlighted as being between 1~and~2\,Myr before $t=$ ``Now'' \citep{WuBik:2016aa}. 
Similarly, the presence of UC\,\ion{H}{ii} regions in W49A indicates a more recent episode of star formation (a few $\times10^5$\,yr ago
or even more recent; \citealt{KawamuraMasson:1998aa,WoodChurchwell:1989aa,MottramHoare:2011aa}). 
The formation period of the UC\,\ion{H}{ii} regions is marked as a time span of $3\times10^5$\,yr before the current time in Fig.~\ref{fig:6}.

\begin{figure*}
  \includegraphics[width=0.99\textwidth]{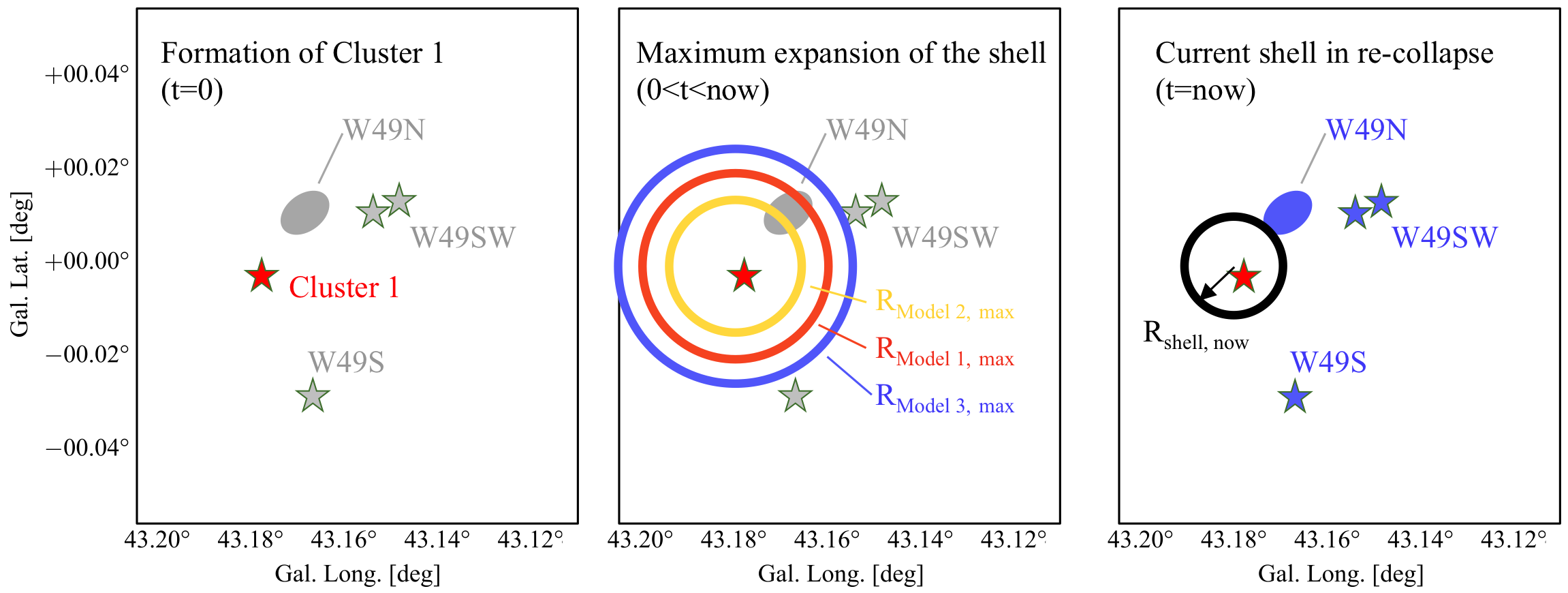}
  \caption{Sketch of the evolution of the feedback-driven shell around Cluster~1 in W49A (see text for details). At $t=0$, Cluster~1 ({\it red}) is formed in the models ({\it left panel}). The feedback from Cluster~1 drives a shell into the molecular gas of W49A ({\it middle panel}). The radii of maximum extension from Fig.~\ref{fig:6} are highlighted (Model 1, 2, 3 in {\it red}, {\it yellow} and {\it blue}, respectively). The shell then re-collapses to its observed extent ({\it right panel}). The other stellar sub-clusters, as well as the Welch-Ring, are highlighted in blue in the right panel to indicate their relative position. In the  left and middle panels, these objects are highlighted in gray to indicate that, depending on the model, they may not have formed yet and that their formation may have been triggered by the feedback-driven shell around Cluster~1. The stellar clusters/Welch ring are not to scale.}
  \label{fig:cartoon}
\end{figure*}

All three models discussed here show the same qualitative evolution: 
A shell, triggered by the formation of a star cluster ({\it left panel} of Fig.~\ref{fig:cartoon}), 
expands to a certain radius inside the molecular cloud ({\it middle panel} of Fig.~\ref{fig:cartoon}), 
at which point the self-gravity of the shell and the gravity of the star cluster start to dominate the force balance. 
The shell would re-collapse to form a new star cluster. Using the observed shell radius and star cluster age as constraints, we infer that the shell is either at the end of re-collapse or at the beginning of a new expansion ({\it right panel} of Fig.~\ref{fig:cartoon}). 

All models predict that Cluster 1 is not powerful enough to disperse the molecular cloud. 
Rather the dispersal would only happen after repeated collapse, which would trigger more star formation. 
Additionally, feedback-triggered star formation in the outskirts of W49A (i.e., fragmentation of the shell or shell collision with outer dense clumps) could increase the total amount of stellar feedback which might eventually become sufficient enough to disperse the cloud.

As can be seen from Models 1-3 in Fig.~\ref{fig:6}, the shell expands farther into surroundings with lower densities, and will also take a longer time to re-collapse for models with lower initial cloud densities. For a given stellar cluster mass with the same strength of feedback, this trend is expected, since the shell has piled up more material at a given radius in a denser medium. 
At even lower densities, feedback from the first cluster would be sufficient to destroy the cloud. However, such densities would be lower than the range of cloud densities estimated from observations.
Similarly, increasing the stellar cluster mass (within the masses probed here) allows the model to push to larger radii, while the timescales for re-collapse are not strongly affected.
More massive stellar clusters provide stronger feedback forces, which accelerate the shell to higher velocities and therefore cause the larger expansion. 
This is shown in Model 4 in Fig.~\ref{fig:6}, for $M_* = 2\times10^4\,{\rm M_\odot}$ at a constant density of $n = 4\times10^3\,{\rm cm^{-3}}$. 

The radius of maximum expansion varies among the different models between the currently observed radius (gray-shaded region B in Fig.~\ref{fig:6}; Model 2) and the projected radius of W49A (gray-shaded region C; Model 3). 
Radial, monotonically decreasing density profiles would imply denser material in the center of W49A. 
The acceleration of the shell would be lower within the maximum expansion radii of the models considered here, since the material which is swept up in the shell grows faster with radius. 
Also, feedback due to thermal pressure in phase 1 (see above) would be lower, as dense gas cools faster. 
Both effects would lead to smaller radii of maximum expansion. 

For all densities probed here, the models confirm that shells driven by Cluster 1 may have at least affected the Welch ring and possibly triggered star formation there \citep[in agreement with][]{AlvesHomeier:2003aa,PengWyrowski:2010aa}. 
They have in common that the feedback-driven shell underwent re-collapse.
On the lower end of the density range justified by the observations, shells have pushed to the outskirts of W49A in the past, and may have affected star formation all across W49A.

\section{Discussion} \label{sec:discussion}
\subsection{Dynamics of re-collapse in W49A} \label{sec:discussion:dynamics}
The stellar feedback models predict a shell in contraction, which is about to induce or may have already induced a new star-formation event. 
From an observational perspective, the double-peaked emission of molecular gas towards Pos.~2 (Cluster~1) is consistent with the presence of a moving shell. 
Both peaks are $\sim$4.5~\kms\ offset from the systemic velocity of $\varv_{\rm LSR} = 8.6$\,\kms \citep{QuirezaRood:2006ab}. 
This indicates expansion or contraction of $\sim$4.5~\kms, if the star cluster, which drives the shell, is at the systemic velocity. 
It is not possible to determine from CO and RRL emission alone, whether the shell is expanding or collapsing. 
The RRL emission appears to be associated more with the high-velocity peak. 
However, the exact peak position is uncertain, due to the low signal-to-noise ratio of the RRL emission at this position. 
Depending on the rest velocity of the star cluster, it may also be possible that expansion or collapse are asymmetric and that one part of the shell is at rest. 

Anti-correlation between molecular and ionized emission in velocity is seen towards Pos.~1 and Pos.~3, offset from Cluster~1. 
At the peak velocities of the ionized gas emission in both positions, the strength of molecular gas emission is weaker in comparison to emission at other velocities. 
If both gas phases at a given velocity are spatially connected, this emission may be weaker as the molecular gas is destroyed. 

The emission structure at both positions can also be affected by local star formation and does not have to be dominated by the shell dynamics alone. 
Position~3 harbors the low-cm continuum emission source ``JJ'', while Pos.~1 contains the sources ``M'' and ``L'' 
\citep[see Fig.~\ref{fig:1}; all three sources are characterized with low-cm continuum emission in][]{De-PreeMehringer:1997aa}.
Position~1 is also located at the edge of the Welch-Ring which harbors multiple UC\,\ion{H}{ii} regions and is dynamically complex \citep[see e.g.,][]{SerabynGuesten:1993aa,WilliamsDickel:2004aa}.
The velocities from the high-order RRLs presented here are similar to the velocities of the individual UC\,\ion{H}{ii} regions (see Sect.~\ref{sec:results:vpeak}), which are determined at significantly higher angular resolution ($<1$\arcsec) from the H92$\alpha$ RRLs \citep{De-PreeMehringer:1997aa}.
The velocity signature therefore may not only be shaped by large-scale motions, such as the expanding bubble, but also by the small-scale dynamics, which are spatially unresolved in the RRL observations presented in this work.

The precise time of the formation of the second generation of stars postulated by the models is difficult to constrain 
since the shell is either close to the end of the re-collapse or at the beginning of the second expansion phase. 
Towards the center of Cluster~1, we do not see evidence of UC\,\ion{H}{ii} regions, 
many young stars or enhanced dust opacities (compared to the rest of W49A), 
which would be indicative of new stars having been formed there recently. 
However, high foreground extinction towards W49A ($A_V>30$\,mag; \citealt{AlvesHomeier:2003aa}) and 
high uncertainties in stellar ages make it impossible to rule out that a second generation of stars has already formed at this position.

Furthermore, as W49A is clearly not spherically symmetric, it is also possible that Cluster~1 did not form at the center of mass and the shell may collapse to a different center.
In the idealized case of a 1D geometry the cloud is forced to re-collapse onto the same cluster.
Depending on the stellar mass distribution, which is intrinsically difficult to measure for the region, 
the re-collapsing gas may fall more towards W49N, which is $\sim$3\,pc offset from Cluster 1 in projection \citep{AlvesHomeier:2003aa} and 
therefore may be within the maximum radius of expansion for large parts of the range of densities in agreement with the observations. 
The youngest clusters dominate feedback. So long as the shell expands to a radius which encompasses both populations the models will be valid in the future. 
However, without clear evidence of the presence or absence of a new population of stars at the center of the shell, 
the molecular cloud is either shortly before, after, or even within a new episode of star formation.

Most likely, W49A is currently in the phase of the formation of this second star cluster. 
Indication of this is the large number of compact and UC\,\ion{H}{ii} regions surrounding Cluster~1, including the Welch-Ring. 
These may have been triggered during the re-collapse of the shell. 
Since some of them are extended (i.e., source "L"; \citealt{De-PreeMehringer:1997aa}), they may already be contributing to the feedback of the bubble.  
These sources may provide the feedback to halt the re-collapse of the shell in the future, or may have already started the second expansion period of such a shell. 

\subsection{Effect of feedback on the molecular gas in W49A} \label{sec:discussion:triggering}
The {\sc warpfield} models were used in Sect.~\ref{sec:results:models} to characterize the expansion of the shell radius. This section discusses to what extent other parts of W49A were affected by feedback and where this feedback may have triggered star formation, for example 
via triggering of clumps of molecular gas into gravitational collapse \citep[e.g.,][]{ElmegreenLada:1977aa,WhitworthBhattal:1994ab,PreibischBrown:2002aa,PreibischZinnecker:2007aa}. 

For all models highlighted in Fig.~\ref{fig:6}, the expanding shell remains confined to the inside of W49A (<6\,pc). 
The radius of maximum expansion is between 2.5 and 5 pc. 
This implies that feedback of Cluster 1 has affected molecular gas in W49N. 
The Welch-ring harbors UC\,\ion{H}{ii} regions, which have an approximate lifetime of $3\times10^5$\,yr \citep{MottramHoare:2011aa}. 
Therefore, these need to have formed after the O/B stars in Cluster 1. 
Feedback may have triggered the star formation in the Welch ring, either at the point of maximum expansion (Model 2; $n\sim7.5\times10^3\,{\rm cm}^{-3}$), or during contraction at re-collapse (Models 1,3,4; $n\lesssim 7.5\times10^3\,{\rm cm}^{-3}$). 
Also, as discussed in Sect.~\ref{sec:discussion:dynamics}, these UC\,\ion{H}{ii} regions may already be part of the second episode of star formation after re-collapse. 

Towards the lower bound of the considered range of densities (Model 3; $n\approx2.5\times10^3\,{\rm cm}^{-3}$), feedback could have pushed to the outskirts of the cloud. 
During the expansion, it may have triggered star formation in molecular clumps in the swept-up molecular gas, which could be the reason for the formation of O stars and UC\,\ion{H}{ii} regions outside of Cluster~1, as well as larger star forming sites such as W49S and W49SW (Fig.~\ref{fig:0}). 
Shell fragmentation could be a possible cause for these clumps to be decoupled from the bulk collapse of the shell. 
This may occur especially when the expansion of the shell has considerably slowed down \citep[e.g.,][ Eq. 14 therein]{McCrayKafatos:1987aa}. 
This is expected at the expansion maximum and could explain why we observe them today at projected distances from Cluster~1 of $\sim5-6$\,pc. 

We note that for Models 1 and 2, the molecular clumps W49S or W49SW at the edge of W49A, 
would not have been affected by the star formation in Cluster 1; they were of independent origin. 
Their formation may be connected to global motions, like cloud-cloud collisions \citep{SerabynGuesten:1993aa} or fragmentation of filamentary inflow \citep{Galvan-MadridLiu:2013aa}. 
This may fit in the picture of sub-clustered star formation, which does not require a causal connection between individual events of star formation in W49A \citep[e.g.,][]{AlvesHomeier:2003aa}.

It must be stated that the models do not predict any ejecta at distances $d>12\,{\rm pc}$, as discussed by \citet{PengWyrowski:2010aa}.
In the picture of re-collapsing shell models, these would need to be formed independently, at least if feedback were driven only by Cluster 1. 
In principle, strong inhomogeneities in the molecular cloud could have allowed feedback to channel out of the surrounding molecular cloud. 
However, as mentioned by \citet{PengWyrowski:2010aa}, high energies would be necessary to drive these ejecta (few $\times10^{50}\,{\rm erg}$).

The previous discussion highlights that the expansion radius strongly depends on the density structure. 
For the models in Sect.~\ref{sec:models}, we assumed a homogeneous density profile. 
If the density instead increased towards the molecular cloud center, the evolution of the shell would have a smaller radius of maximum expansion (see Sect.~\ref{sec:results:models}). 
However, the models would still predict a re-collapsing shell similar to Models~1~and~2.
Since most of the expansion occurs in the momentum-driven phase (see Sect.~\ref{sec:results:initialconditions}), an inhomogeneous density distribution may affect the expansion of the shell locally. 
The expansion of the shell would necessarily be asymmetric, that is, expanding further in directions of lower densities and less in directions of higher densities as compared to the expansion in the case of an average, homogeneous density distribution as delineated here. 
Modeling the exact shape and the precise impact on the cloud of the feedback-driven shell requires 3D simulations, as further discussed in Sect.~\ref{sec:discussion:future}, and requires us to make assumptions regarding the initial density distribution of the cloud. 
Keeping in mind the limitation of such an analysis, the models presented here give an estimation of the evolution of the feedback-driven shell using the simplest assumption on a density profile, namely that it is homogeneous. 

To summarize, the {\sc warpfield} models show that the feedback shell has affected different subsections of the molecular cloud, depending on the assumed density distribution of the molecular cloud.
The compression of gas by the feedback-driven shell potentially triggers gravitational collapse of dense gas clumps and the formation of new stars. 
While widespread star formation is observed around Cluster~1, the effectiveness of triggering remains disputed in the literature \citep[e.g.,][]{DaleBonnell:2011aa,KrumholzBate:2014aa}, and is yet to be demonstrated to alter the star formation efficiency of molecular clouds in general and of W49A in particular. 
Nonetheless, we speculate that this can constitute an additional hypothesis for the high star formation activity at least of parts of W49A. 
Star formation in regions of the cloud that were not affected by feedback needs to have occurred in different ways. 
Therefore, these models do not necessarily rule out other hypotheses on the formation scenario of stars in W49A, such as sub-clustered star formation or cloud-cloud collisions, which have already been discussed in the literature. 

\subsection{Star formation rate and multiple generations of stars} \label{sec:discussion:sfe}
Re-collapsing shells predict episodic events of star formation in a molecular cloud. 
While the previous section discussed potential triggering of star formation during the expansion of the shell into the molecular cloud of W49A, 
here we focus on the formation of a new star cluster when the shell re-collapses. 
At this time, the accumulated molecular gas of the shell-material is concentrated inside a small radius and forms a new star cluster \citep{RahnerPellegrini:2017aa,RahnerPellegrini:2018aa}.

At each re-collapse event, stars are formed at a certain instantaneous star formation efficiency (SFE), 
\begin{equation}
\epsilon = M_*/M_{\rm cloud}, 
\end{equation}
with $M_*$ the mass of the star cluster originating from a single star formation event, 
and $M_{\rm cloud}$ the cloud mass prior to the formation of a cluster \citep[e.g.,][]{Murray:2011aa}. 
If star clusters continue to form with a fixed SFE for each star formation event (we adopt values of $\epsilon = 0.05$ and 0.09), 
it would take multiple re-collapse events (each of which leads to a new star formation event) to disperse the cloud. 
Since the main focus of the model comparison was to follow the first expansion cycle of the feedback shell from Cluster~1, 
we assume here for simplicity that the SFE at future collapse events is the same as at the initial cluster formation.
Changes in the SFE, however, are expected (as, e.g., in 30 Dor; \citealt{RahnerPellegrini:2018aa}), 
since the SFE depends on the physical conditions of the re-collapsing gas, 
which are likely to be different from those during the first episode of star formation.
The recently formed stars will eventually power a common feedback shell together with the already existing star cluster, which will affect the evolution of this shell and change the time-scale between re-collapse events.

The re-collapse events have implications for the age distribution of the future star cluster.
Different generations of stars would be expected in W49A, induced by recurring collapse events and the subsequent formation of a new star cluster. 
Their age difference in the case of W49A will, however, be small compared to other star-forming regions, such as 30 Doradus (see Sect.~\ref{sec:discussion:models}).
The models for Cluster~1 predict an age difference of only 1-2\,Myr. 
The wide-spread star formation in W49A suggests that stars may also form in between (see Sect.~\ref{sec:discussion:triggering}). 
The age differences are therefore likely to be lower than 1\,Myr, 
which is well within typical uncertainties of the ages derived for O stars, 
and therefore intrinsically difficult to differentiate observationally.

\subsection{W49A in comparison to the massive star-forming region 30 Doradus} \label{sec:discussion:models}
One of the most prominent prototypes for an extreme high-mass star-forming region can be found in 30 Doradus, which is located in the LMC.
It has the observational advantage of being visible in the optical, which is not possible for W49A due to the obscuration of the dust by the Milky Way. 
There are obvious differences between both regions in molecular gas mass (depending on method and radius: $10^{6-7}\,{\rm M_\odot}$ for 30~Doradus; e.g., \citealt[][]{DobashiBernard:2008aa,Faulkner:1967aa,SokalJohnson:2015aa}; vs. $10^{5.3-6}$ for W49A; e.g., \citealt[][]{Galvan-MadridLiu:2013aa}) and stellar cluster mass ($7\times10^4-5\times10^5$\,${\rm M}_\odot$ for 30~Doradus; \citealt[][]{SelmanMelnick:1999aa,BoschSelman:2001aa,BoschTerlevich:2009aa,CignoniSabbi:2015aa}; vs. $5-7\times10^4$\,${\rm M}_\odot$ for W49A; \citealt{HomeierAlves:2005aa}). 
Here, we focus the discussion on feedback-related star formation in both star-forming regions. 

The star cluster NGC 2070 in 30 Doradus contains a sub-cluster, R136, which is younger ($\sim 1\,{\rm Myr}$) than the older stellar population of NGC 2070 ($\sim 5\,{\rm Myr}$; \citealt{BrandlSams:1996aa,WalbornBlades:1997aa,MasseyHunter:1998aa,SelmanMelnick:1999aa,SabbiLennon:2012aa,CignoniSabbi:2015aa,CrowtherCaballero-Nieves:2016aa}). 
\citet{RahnerPellegrini:2018aa} found that the formation of R136 can be naturally connected to the formation of NGC 2070 by the re-collapse 
of an expanding shell from the first star formation event, and the subsequent formation of R136. 
While the stars in W49A are of similar age to R136, there is no distinct older population. 
Rather, there seems to be only an even younger population of UC\,\ion{H}{ii} regions present in W49A, with an age difference to the O-star population of 1-2\,Myr (see Sect.~\ref{sec:discussion:triggering}). 
The model confirms this difference in star formation history:
while for 30 Dor, the models find re-collapse events 3-4 Myr after the initial event of star formation, 
the re-collapse is predicted earlier for W49A, within 2\,Myr after initial star formation.

The larger re-collapse interval in 30 Doradus is due to a combination of different cloud density, stellar cluster mass, and metallicity.
The models require higher initial molecular gas densities in W49A to satisfy the observational constraints ($n\gtrsim2.5\times10^3\,{\rm cm}^{-3}$) than have been found for 30 Doradus ($n\gtrsim5\times10^2\,{\rm cm}^{-3}$). The mass of the first cluster of stars is a factor of a few higher in 30 Doradus and its metallicity is lower at $Z\approx0.5\,{\rm Z_\odot}$ \citep{LebouteillerBernard-Salas:2008aa,ChoudhurySubramaniam:2016aa}.

The lower metallicity in 30 Doradus changes the shell evolution in a nontrivial manner.
While \citet{LopezKrumholz:2011aa} suggest weaker winds in low-Z systems, \citet{RahnerPellegrini:2017aa} finds that the longer cooling time can lead to winds being more important, as the expansion of the shell stays in phase 1, the energy-driven phase, for a longer time.
\citet{PellegriniBaldwin:2011aa} finds that the shell structure is over-pressured near the cluster by X-ray bubbles over radiation pressure, consistent with the results by Rahner et al., and in opposition to \citet{LopezKrumholz:2011aa}. 
In turn, accelerations of the shell in phase 2, the momentum-driven phase, may decrease. 
Low metallicities imply weaker stellar wind luminosities (e.g., \citealt{LopezKrumholz:2011aa}) and affect the coupling of radiation to the shell by changing the shell structure \citep{RahnerPellegrini:2017aa}: 
The confining pressure from the winds of the shell is weaker, which leads to lower densities in the shell \citep[see][eq. 14]{RahnerPellegrini:2017aa}. 
In turn, these lower densities result in a decrease in radiation pressure on a given part of the shell, as it absorbs less ionizing radiation than a high density shell, 
because the absorption of ionizing radiation by hydrogen depends on the recombination rate and is proportional to $n^2$. 
The effect of metallicity is therefore complicated and depends on the details of the investigated model whether there is faster or slower expansion.
Further, \citet{PellegriniBaldwin:2011aa} found significant fractions of the 30 Doradus region to be dominated by optically thin (fully ionized) gas  leading to a decoupling between radiation and the gas, a result that can only be explained by fully accounting for feedback and ISM coupling and cooling, and not from simple scaling relations.

These models provide a framework with which to understand how feedback (both positive and negative) regulates star formation and the associated timescales. 
The fact that models with re-collapse of a feedback shell are compatible with observations in W49A 
indicates that re-collapse may not be a phenomenon that is unique to 30 Doradus. 
This would imply that feedback is not only responsible for cloud dispersal, 
but could also be the means by which star formation  regulates the rate of stellar birth by introducing a timescale for subsequent cluster-formation events.
The timescale would be much shorter (<2 Myr) in W49A than in 30 Doradus, and consequently it would also be harder to observe (see Sect.~\ref{sec:discussion:sfe}). 
Other clusters besides 30 Doradus and potentially W49A may be affected by feedback-regulated star formation episodes. 
According to \citet{RahnerPellegrini:2018aa}, these need to have a SFE of $\lesssim10\%$ and average densities of $n\ge5\times10^2\,{\rm cm^{-3}}$. 
Investigation of clusters that meet these criteria is necessary to decipher whether or not  this is a common scenario in the formation of massive star clusters.

\subsection{Future investigations}\label{sec:discussion:future}
Detailed observations are needed of the morphology of the molecular gas and the stellar content as seen in infrared wavelengths in order to shed more light on the physical processes dominating the dynamics and evolution of W49A. 
While much work has been done on W49N, more studies focussing on the connection of the large-scale dynamics inside and outside of the region are necessary.
With the Atacama Large Millimeter/sub-millimeter Array (ALMA\footnote{\url{http://www.almaobservatory.org}}), kinematics of molecular and ionized gas can be mapped at significantly higher resolution than the observations of ionized and molecular gas applied here. 
Due to the high extinction towards W49A of $A_V>30$\,mag \citep{AlvesHomeier:2003aa}, soft, diffuse X-ray emission from wind bubbles \citep{TownsleyFeigelson:2003aa,TownsleyBroos:2014aa,TownsleyBroos:2018aa} are difficult to observe. 
Chandra observations of hard X-rays for W49A exist, and have been discussed for W49N, including both wind-driven bubbles and other possible formation scenarios \citep{TsujimotoHosokawa:2006aa}. 
An interpretation of these in the light of wind-driven feedback is therefore nontrivial and remains to be explored for Cluster 1. 
The James Webb Space Telescope (JWST\footnote{\url{https://www.jwst.nasa.gov}}) will yield high-sensitivity imaging of many star-forming regions in the Galaxy. 
This will provide better insight into the most embedded populations of O stars in regions like W49A. 

The 1D models from \citet{RahnerPellegrini:2017aa} are well suited for exploring a large parameter space in molecular cloud mass, star formation efficiency, and density. 
While these models contain all the necessary physics to describe the expanding shell, modeling a shell in three spatial dimensions is necessary. 
Three-dimensional models can better account for escape routes for feedback in the energy-driven phase at the beginning of the expansion, 
as well as for instabilities in the shell, which may alter the effectiveness of feedback \citep{DaleNgoumou:2014aa}, even though the impact of the shell structure on the escape fraction of ionizing radiation is accounted for in the 1D models. 
More importantly, it is the density structure that has the greatest effect on  the geometry and the timescales of the expansion. 
As seen in W49A, many star-forming regions appear not to be spherically symmetric, and expansion will not occur isotropically into a uniform medium. 
These models are naturally more computationally intensive.
Such studies have been carried out \citep[e.g.,][]{HowardPudritz:2016aa,PetersBanerjee:2010aa}, although these have not yet included all the relevant physics. 
However, given that with the 1D models we can explore the parameter spaces very well, the results from this work may provide the average initial cloud parameters for more detailed modeling. 

\section{Conclusions} \label{sec:conclusion}
The study presented here investigates hydrogen radio recombination line emission of the H151$\alpha$-H156$\alpha$ and H158$\alpha$ transitions at frequencies between 1.6 and 1.9 GHz with the VLA in C-configuration towards W49A from the THOR survey \citep{BeutherBihr:2016aa}, as well as archival CO observations from the CHIMPS and GRS surveys \citep{RigbyMoore:2016aa,JacksonRathborne:2006aa}.
We find shell-like RRL emission around the infrared star cluster at the center of W49A, Cluster~1. 
We find double peaked $^{13}$CO(3-2) emission towards Cluster~1, again indicative of emission from shell-like geometries. 
The RRL emission is rather broad over all velocities between the two CO components, but peaking towards the redshifted one. 
Towards positions at the edges of the shell-like bubble, the velocities of the ionized and molecular gas emission are anti-correlated. 
This anti-correlation may be a signature of interaction of the ionized shell with the molecular envelope. 

\smallskip
\noindent
One-dimensional models of expanding shells around massive star clusters \citep[{\sc warpfield},][]{RahnerPellegrini:2017aa} are used to investigate the evolution of the shell-like bubble in W49A. 
Given the observational initial conditions of stellar cluster mass, age of the O-star population, and the current radius of the shell, all models predict re-collapse of the shell after the first star formation event. 
Feedback of the first formed cluster is therefore not strong enough to disperse the cloud in the first instance. 

\smallskip
\noindent
The evolution of the shell however strongly depends on the assumed densities. 
In some of the models, the shell has expanded to the outskirts of W49A. 
With this, it could have affected molecular gas in the entire region of W49A.
In this case, a causal connection between feedback from cluster 1 and star formation all across W49A is possible. 
However, for most of the models in the range of the assumed densities, the feedback shell did not expand to the outskirts of W49A. 
Hence, it is more likely that only limited parts of W49A were affected by feedback from the central stellar cluster, while stars in the outer parts of W49A formed independently. 
To what extent this feedback altered the physical conditions in the surroundings and nurtured (i.e., triggered) star formation, is left to future studies. 

\smallskip
\noindent
This modeling presents another alternative to sub-clustered star formation and cloud-cloud collision models, without ruling out either. 
In comparison to the star-forming region 30 Doradus, indications for re-collapse are less clear due to difficulties in identifying two distinct stellar populations. 
On the other hand, as W49A is younger, it is likely to continue to form stars, possibly in an episodic mode. 

\begin{acknowledgements}
We would like to thank the referee for the detailed, helpful, and insightful comments, which considerably improved the paper.
M.R.R. is a fellow of the International Max Planck Research School for Astronomy and Cosmic Physics (IMPRS) at the University of Heidelberg.
H.B., M.R.R., Y.W., J.S. and J.C.M. acknowledge support from the European Research Council under the Horizon 2020 Framework Program via the ERC Consolidator Grant CSF-648505.
\mbox{M.R.R., D.R., H.B., E.W.P, S.C.O.G. and R.S.K.} acknowledge support from the Deutsche Forschungsgemeinschaft (DFG) via Sonderforschungsbereich (SFB) 881 ``The Milky Way System'' (sub-projects B1, B2 and B8). 
S.C.O.G., E.W.P. and R.S.K. further acknowledges support from the DFG via Priority Program SPP 1573 ``Physics of the Interstellar Medium'' (grant numbers KL 1358/18.1, KL 1358/19.2, and GL 668/2-1) and from the European Research Council via the ERC Advanced Grant STARLIGHT (project number 339177).
The research was carried out in part at the Jet Propulsion Laboratory, which is operated for NASA by the California Institute of Technology.
R.J.S. acknowledges support from an STFC Ernest Rutherford fellowship.
S.E.R. acknowledges support from the European Union's Horizon 2020 research and innovation programme under the Marie Sk{\l}odowska-Curie grant agreement \mbox{\# 706390}.
F.B. acknowledges funding from the European Union's Horizon 2020 research and innovation programme (grant agreement \mbox{No 726384} - EMPIRE).

\end{acknowledgements}
\bibliographystyle{aa}
\bibliography{34068corr}

\appendix
\section{Additional plots}

\begin{figure*}
  \includegraphics[width=0.99\textwidth]{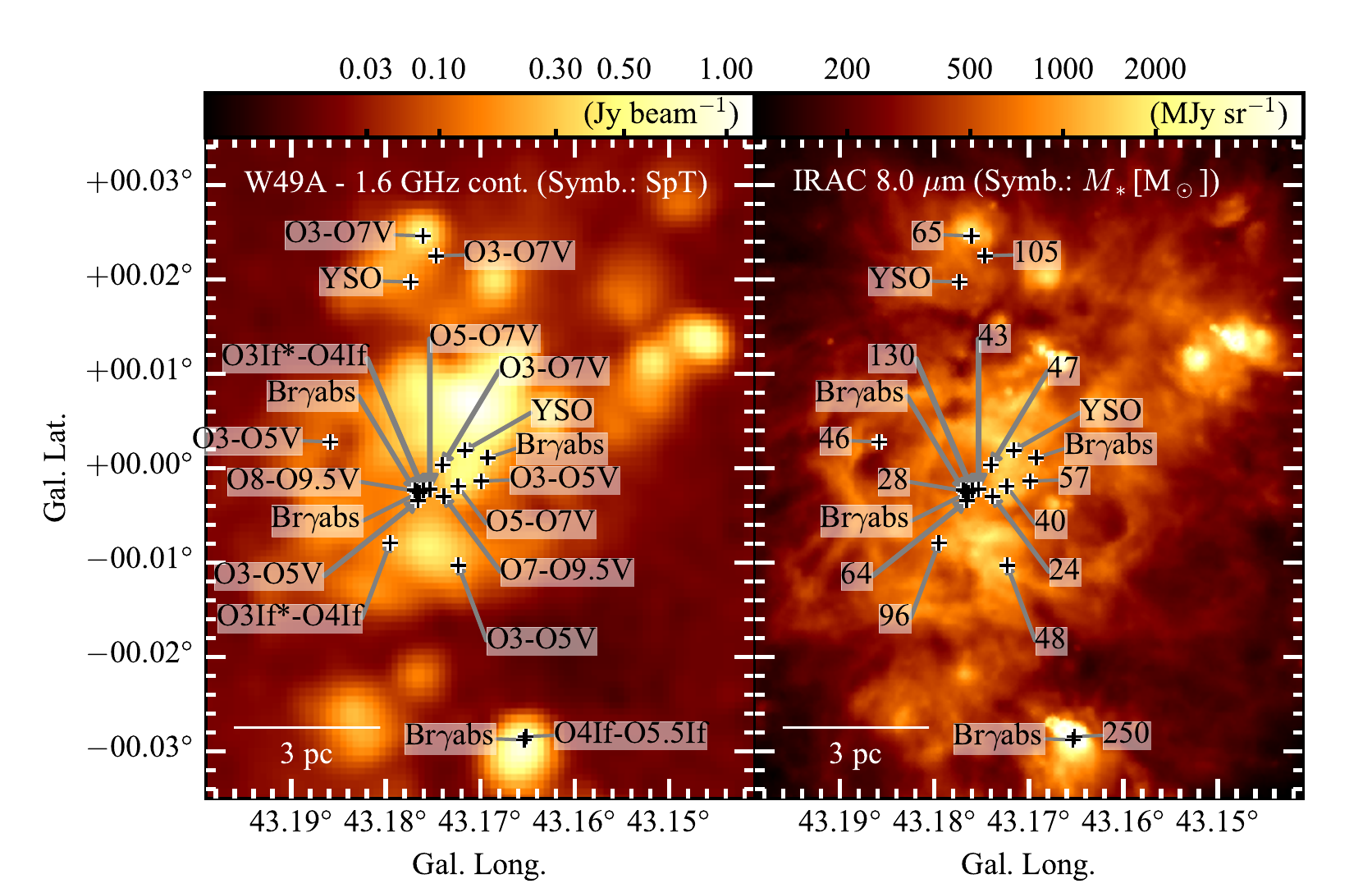}
  \caption{Spectral types (SpT) and masses of O stars in W49A, as determined in \citet{WuBik:2016aa}. 
                Symbols and data are as in Fig.~\ref{fig:0}. Both 1.6 GHz continuum emission ({\it left}; THOR; \citealt{WangBihr:2018aa}) 
                and GLIMPSE Spitzer IRAC 8.0 $\mu$m emission ({\it right}; \citealt{BenjaminChurchwell:2003aa,ChurchwellBabler:2009aa}) are shown in color-scale.}
  \label{fig:a3}
\end{figure*}

\begin{figure*}
  \includegraphics[width=0.99\textwidth]{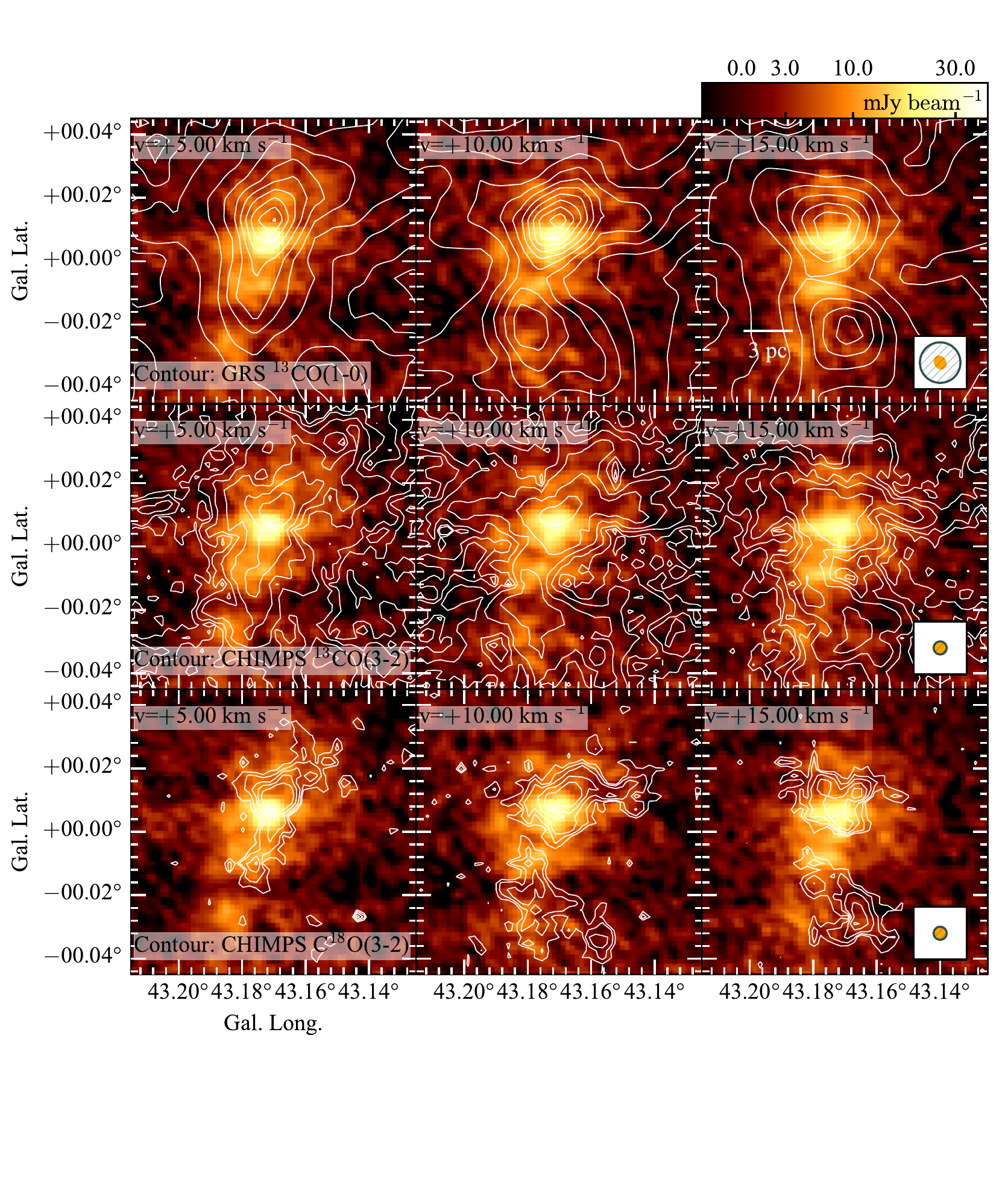}
  \caption{RRL emission as in Fig.~\ref{fig:3} for 5-15\,\kms. Overlaid as contours are GRS $^{13}$CO(1-0), CHIMPS $^{13}$CO(3-2) and
    C$^{18}$O(3-2). The contours are shown for the GRS $^{13}$CO(1-0) emission at levels of 
    0.075, 0.125, 0.25, 0.5, 1.0, 1.5, 2.0, 2.5, 3.0, 3.5, 4.0, 4.5, and 5.0 K, for the
    CHIMPS $^{13}$CO(3-2) emission at levels of 0.45, 0.9, 1.79, 3.57, 7.13, and 14.23 K, and for the 
    C$^{18}$O(3-2) emission at levels of 0.65, 1.03, 1.63, 2.59, and 4.1 K. 
    The angular resolution of the data is shown in the  right column. 
    The colored beam indicates the resolution of the RRL data ($16\farcs8\times13\farcs8$), 
    with the beam of the CO emission overlayed in gray 
    (46\arcsec\ for GRS $^{13}$CO(1-0) emission and 15\arcsec\ for CHIMPS $^{13}$CO(3-2) and C$^{18}$O(3-2) emission).}
  \label{fig:4}
\end{figure*}

\end{document}